\documentclass[letterpaper,floatfix,reprint,superscriptaddress,footinbib,pre,aps]{revtex4-1}
\usepackage[utf8]{inputenc}
\usepackage[english]{babel}
\usepackage{microtype}

\usepackage{amsmath}
\usepackage{amsfonts}
\usepackage{amssymb}
\usepackage{bm}
\usepackage{enumitem}

\usepackage[hyperindex,colorlinks,hyperfootnotes = false,citecolor = blue]{hyperref}
\usepackage[usenames,dvipsnames,svgnames,table]{xcolor}

\usepackage{cleveref}
\usepackage{graphicx}

\begin{document}
\title{Multiscale dynamical embeddings of complex networks}
\date{\today}

\author{Michael T. Schaub}
\email{mschaub@mit.edu}
\affiliation{Institute for Data, Systems and Society, Massachusetts Institute of Technology, Cambridge, MA 02139, USA}
\affiliation{Department of Engineering Science, University of Oxford, Oxford, UK}
\author{Jean-Charles Delvenne}
\affiliation{ICTEAM, Universit\'e catholique de Louvain,  B-1348 Louvain-la-Neuve, Belgium} 
\affiliation{CORE, Universit\'e catholique de Louvain,  B-1348 Louvain-la-Neuve, Belgium} 
\author{Renaud Lambiotte}
\affiliation{Mathematical Institute, University of Oxford, Oxford, UK }
\author{Mauricio Barahona}
\email{m.barahona@imperial.ac.uk}
\affiliation{Department of Mathematics, Imperial College London, London SW7 2AZ, UK}

\begin{abstract}
Complex systems and relational data are often abstracted as dynamical processes on networks. 
To understand, predict and control their behavior, a crucial step is to extract reduced descriptions of such networks.
Inspired by notions from Control Theory, we propose a time-dependent dynamical similarity measure between nodes, which quantifies the effect a node-input has on the network.
This dynamical similarity induces an embedding that can be employed for several analysis tasks.
Here we focus on (i)~dimensionality reduction, i.e., projecting nodes onto a low dimensional space that captures dynamic similarity at different time scales, and (ii)~how to exploit our embeddings to uncover functional modules.
We exemplify our ideas through case studies focusing on directed networks without strong connectivity, and signed networks.
We further highlight how certain ideas from community detection can be generalized and linked to Control Theory, by using the here developed dynamical perspective.
\end{abstract}
\maketitle

\section{Introduction}
Complex systems comprising a large number of interacting dynamical elements commonly display a rich repertoire of behaviors across different time and length scales.
Viewed as collections of coupled dynamical entities, the dynamical trajectories of such systems reflect how the topology of the underlying graph constrains and moulds the local dynamics. 
Even for networks without an intrinsically defined dynamics, such as networks derived from relational data, a dynamics is often associated to the network data to serve as a proxy for a process of functional interest, e.g., in the form of a diffusion process.
Comprehending how the network connectivity influences a dynamics is thus a task arising across many different scientific domains~\cite{Newman2010,Arenas2008a,Bullmore2009}.

However, it is often impractical to keep a full description of a dynamics and the network for system analysis. 
In many cases it may be unclear how such an exhaustive description could be interpreted, or whether such finely detailed data is necessary to understand the phenomena of interest.
Accordingly, many studies aim to reduce the complexity of the system by extracting lower dimensional descriptions, which explain the behavior of interest in a simpler manner with fewer, aggregated variables. 

This reductionist paradigm may be illustrated with the process of opinion formation in a social network. 
In general, there will be many actors in the network, organized in different social circles and influenced by various agents, media, etc.
While the full dynamics is highly complex and variable, the globally emerging dynamics may still evolve on an effective subspace of low dimensionality, such that a coarse-grained description at the aggregated level of social circles may be sufficient to describe the process.

\begin{figure*}[tb!]
 \centering
 \includegraphics{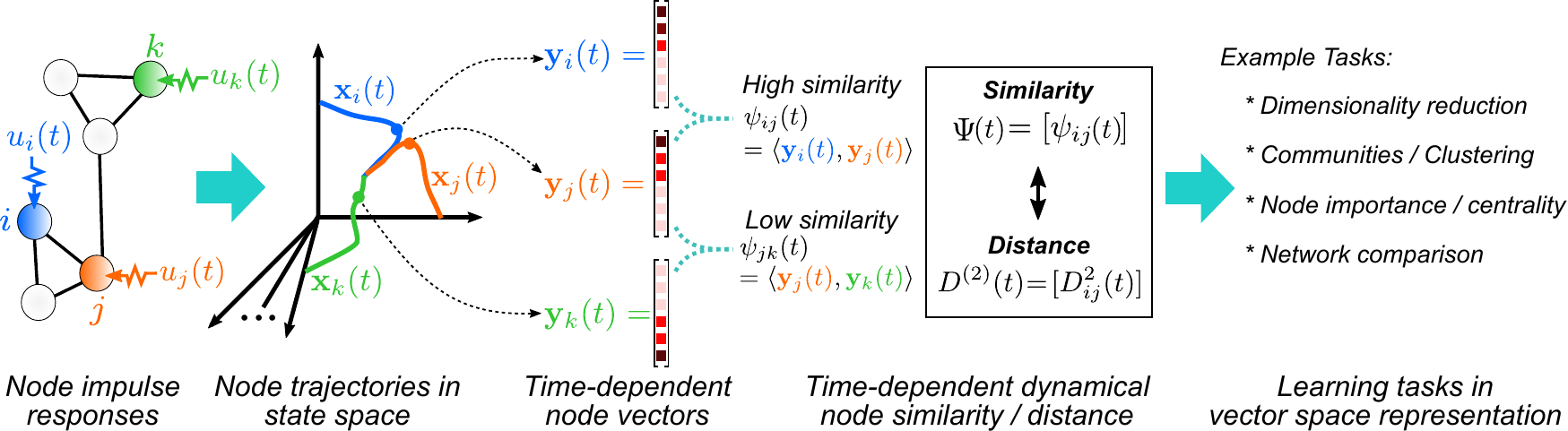}
 \caption{\textbf{Schematic of constructing dynamical similarity measures.} 
     Impulses are applied as inputs to different nodes of the network. 
     The responses in time are interpreted as node vectors evolving in state space, and can be compared, e.g., via an inner product from which we construct the similarity matrix $\Psi(t)$, or alternatively, its associated distance $D^{(2)}$. 
     Nodes that drive the system similarly (differently) within the projected subspace are assigned a high (low) similarity score.
 The thus derived vector space representation of the nodes can be used for a number of different learning tasks.}%
 \label{fig:schematics}
\end{figure*}

A classical source for dimensionality reduction is the presence of symmetries in the system~\cite{Pecora2014,Sorrentino2016}, or the presence of homogeneously connected blocks of nodes.
Yet, \textit{strict, global} symmetries are rare in real complex systems, and while a statistical approach can be used to interpret the irregularities as random fluctuations from an ideal model, e.g., stochastic blockmodels~\cite{Holland1983,Snijders1997}, such models often posit strong locality assumptions such as i.i.d.\ edges.
In particular, many global features, such as cyclic structures and higher order dynamical couplings, cannot be captured within such a block structure paradigm~\cite{Schaub2012,Banisch2015}.

Embedding techniques, which define an (often low-dimensional) representation of the network and its nodes in a metric vector-space, have thus gained prominence recently~\cite{Perozzi2014,Grover2016}, as they allow us to use a plethora of computational techniques that have been developed for analysing data in vector spaces.
Thus far most of these techniques have focussed squarely on representing topological information such as communities, e.g., by using a geometry induced by diffusion processes.
However, networks often come equipped with a more general dynamics than diffusion, or contain signed and directed edges, for which it is not clear how to define an appropriate diffusion.

Inspired by notions from control theory, here we propose a dynamical embedding of networks that can account for such cases. 
Our embedding associates to each node the trajectory of its (zero-state) impulse response.
As illustrated in Figure~\ref{fig:schematics} we construct, for each time $t$, a representation of the nodes in signal (vector)-space, which provides us with a dynamics-based, geometric representation of the system, and associated similarity and distance measures.
Nodes that are close in this embedding  induce a similar state in the network at a particular time scale $t$ following the application of an impulse. 

We can exploit this vector space representation and the associated similarity and dual distance measures for various analysis tasks. 
While such representations are amenable for general learning tasks, in this work we focus on two examples that highlight particular features of interest for dynamical network analysis.
First, we illustrate how \emph{low-dimensional} embeddings of the system can be constructed --- providing a dimensionality reduction of the system in continuous space.
Second, we illustrate how these ideas can be exploited to uncover dynamical modules in the system, i.e., groups of nodes that act approximately as a \textit{dynamical unit over a given time scale}, and discuss how these modules can be related to notions from Control Theory --- an important topic that has gained prominence recently in network theory.
We further show how our embeddings provide links between certain ideas from model order reduction and control theory on the one hand, and notions from network analysis and low-dimensional embeddings on the other hand.

The paper is structured as follows.
In \Cref{sec:methods} we introduce dynamical similarity and distance measures as well as their theoretical underpinnings, and discuss various interpretations of the measures we derive.
The similarity measures and the associated distances can be utilized in different ways for system analysis as we illustrate in \Cref{sec:dim_reduction_ranking,sec:SItribes,sec:func_modules}.
\Cref{sec:dim_reduction_ranking} focusses on applications of our embeddings for ranking, as illustrated by the analysis of an academic hiring network.
\Cref{sec:SItribes} highlights how our framework can be used for (dynamical) dimensionality reduction for signed social networks, using a network of tribal interactions as example.
\Cref{sec:func_modules} then discusses how we can detect functional modules in a signed network of neurons.
We conclude with a brief discussion in \Cref{sec:discussion}.

\section{Dynamical embeddings of networks and node distance metrics}\label{sec:methods}

\subsection{An illustrative example of dynamical node similarity}
To fix ideas, let us envision our system in the form of a discrete time random walk dynamics on a network of $n$ nodes: 
\begin{equation}
    \mathbf{y}_{t+1} = M^\top \mathbf y_{t},
\end{equation}
where \(M = K^{-1}A\) is the transition matrix of an unbiased random walker, \(A\) is the (weighted) adjacency matrix, and \(K=\text{diag}(A\bm 1)\) is the diagonal (weighted) out-degree matrix.

The entries of vector $\mathbf{y}(t)$ correspond to the the probabilities of the random walker to be present at each node at time $t$.
As each variable is identified with a node of a graph, we can assess whether two nodes play a similar dynamical role as follows.
Let us inject an impulse at node $i$ at time $t=0$ and observe the response of the system $\mathbf{y}_{i}(t) \in \mathbb{R}^n$. 
In the context of our diffusion system this means fixing all the probability mass at node $i$ at time $t=0$ and observing its temporal evolution over time.
We define the mapping $i \mapsto \mathbf{y}_{i}(t)$, which associates to each node its zero-state impulse response. 
This mapping embeds the nodes into a space of signals, and we can thus use any suitable similarity measure between the signals $\mathbf{y}_{i}(t)$ and $\mathbf{y}_{j}(t)$ to define a \textit{node similarity}. 

To quantify whether the impact of node $i$ in the network is aligned with the impact of node $j$ at a particular time $t$, a wide variety of similarity functions between $\mathbf{y}_i(t)$ are possible, including nonlinear kernels~\cite{Schoelkopf2002}.
However, we find it convenient to use the standard bilinear inner product:
\begin{align}\label{eq:innerproduct}
    \Psi(t) & =  \begin{bmatrix} 
    \psi_{ij}(t)
\end{bmatrix}_{i,j=1,\ldots,n}  \\
        \text{with} \quad \psi_{ij}(t) &= \langle  \mathbf{y}_{i}(t) , \mathbf{y}_{j}(t)  \rangle = \mathbf{y}_{i}{(t)}^\top\mathbf{y}_{j}(t). \nonumber
\end{align}
Note that $\Psi$ does in general \emph{not} indicate the presence of regions in which the flow is trapped; instead, the similarity between two nodes $i,j$ is defined by how aligned the influence of an impulse emanating from nodes $i,j$ is after a time $t$.
Accordingly, a high dynamical similarity does not necessitate direct proximity in the underlying graph.

\begin{figure}[tb!]
 \centering
 \includegraphics{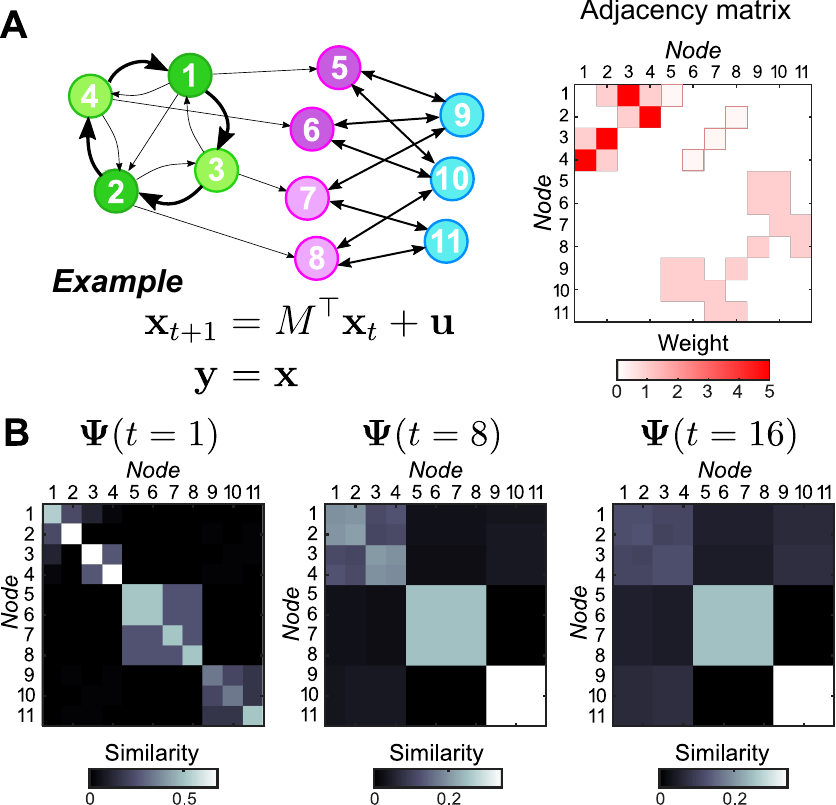}
 \caption{\textbf{Constructing dynamical similarity measures.}  
     \textbf{A} Visualization of an asymmetric directed network (not strongly connected) and its adjacency matrix.
     Note that the which contains a bipartite (disassortative) substructure. 
     \textbf{B} Similarity matrix $\Psi(t) = M^t{[M^t]}^\top$ for times $t=\{1,8,16\}$.}%
     \label{fig:schematics2}
\end{figure}

Figure~\ref{fig:schematics2} illustrates some of the key aspects of the dynamical similarity measure defined in those terms for an example graph equipped with a diffusion dynamics.
As can been seen, e.g., nodes 5 and 6 (in the pink group) behave similarly over short time-scales ($t=1$), whereas the other nodes behave more distinctly. 
Over intermediate time scales ($t=8$), the similarity of the nodes converges into four blocks: the pink group (nodes $5-8$), the cyan group (nodes $9-11$), and two subgroups (nodes $1-2$ and nodes $3-4$) within the green cycle subgraph (nodes $1-4$). 
At longer time scales ($t=16$) the similarity of the nodes, may be approximated by three dynamical blocks (green, pink, cyan).

While the above example hints at how our similarity measure may be employed for the detection of dynamically cohesive modules, note that in contrast to many methods used to detect graph communities based on diffusion~\cite{Delvenne2010,Rosvall2008,Pons2005,DeDomenico2017} or on the propagation of a perturbation~\cite{Arenas2008,Kolchinsky2015}, the above formulation in terms of response dynamics does not require the graph to be strongly connected.
Further, there is also no notion of `assortative' network structure built into the similarity measure: as seen in Figure~\ref{fig:schematics2}, cyclic and bipartite structures are identified in a naturally interpretable manner over particular time scales.
However, we emphasize here that the purpose of the embedding is not to detect topological meaningful communities, but to quantify in how far nodes behave dynamically similar, which is a different objective~\cite{Schaub2017}.

\subsection{General dynamical similarity and distance measures}
Let us now formalize the above ideas in more general terms and consider the following linear dynamics:
\begin{subequations}\label{eq:gen_setup}
\begin{flalign}
    \mathbf{\dot{x}} &= \mathcal A \mathbf{x} + \mathcal B\mathbf{u} \quad  \mathcal{A} \in \mathbb{R}^{m\times m}, \quad \mathcal{B}\in\mathbb{R}^{m\times p}\\
    \mathbf{y} &= \mathcal C\mathbf{x} \qquad  \qquad \mathcal{C} \in \mathbb{R}^{n\times m},
\end{flalign} 
\end{subequations}
where $\mathbf{x} \in \mathbb{R}^{m} , \mathbf{y}  \in \mathbb{R}^{n}, \mathbf{u}  \in \mathbb{R}^{p}$ are the state, the observed state, and the input vectors, respectively. 
While discrete-time systems are also of interest (Fig.~\ref{fig:schematics2}), we will in the following primarily stick to the continuous time formulation for simplicity.
All our results can however be naturally translated to discrete time.


Based on \cref{eq:gen_setup}, we collect the (zero state) impulse responses $\mathbf{y}_{i}$ for every node $i$ and assemble them into the matrix 
$Y(t) = [\mathbf{y}_{1}, \ldots, \mathbf{y}_{n}] = \mathcal{C}\exp(\mathcal{A} t) \mathcal{B}$.
We now define the similarity matrix $\Psi(t)$ as:
\begin{equation}\label{eq:Psi}
    \Psi(t) = Y^\top \mathcal W Y 
    =  \mathcal B^\top \exp{(\mathcal At)}^\top\mathcal{C}^\top \mathcal{W} \, \mathcal{C}\exp(\mathcal A t) \mathcal B. 
\end{equation}
where we allowed for a weighted inner-product by including the matrix $\mathcal{W}$.
For instance, we may choose $\mathcal W$ to correspond to a degree weighting $\mathcal{W} =\text{diag}(\mathbf{d})$, where $\mathbf{d}$ is the vector of node degrees, such that the influence on nodes with a higher-degree will be weighted more strongly.

Instead of an inner product, other measures of similarity between the responses $\mathbf{y}_i$ could be considered, such as different correlations, or information theoretic measures.
However, defining the similarity via an inner product is conceptually appealing as there is an associated distance matrix $D^{(2)}(t)$, whose entries correspond to a squared Euclidean distance of the form: 
\begin{equation}\label{eq:distance}
    D_{ij}^{(2)}(t) = \|\mathcal W^{\frac{1}{2}}\left(\mathbf{y}_{i}(t) \!- \mathbf{y}_{j} (t)\right)\|^2 = \psi_{ii} + \psi_{jj} - 2 \psi_{ij}. 
\end{equation}

The time parameter inherent to both $\Psi(t)$ and $D^{(2)}(t)$ may be understood as a sampling of the network dynamics at a particular time-scale, which enables us to focus on different time scales of interest.
For instance, we can ignore fast paced transients $\tau$ and consider only long-time behaviors for $t \ge \tau$.
As a concrete example, consider again a diffusion dynamics in discrete time as shown in Figure~\ref{fig:schematics2}B-C, where different time-scales provide different meaningful descriptions.
Setting $t=1$ amounts effectively to a structural analysis in which merely the direct coupling is considered; setting $t>1$ amounts to integrating information over multi-step pathways~\cite{Schaub2012,Delvenne2013}.
If we are interested in features persistent over a range of times, we may also integrate over $t$.
This integration eliminates the time-dependence, and we recover an interesting connection to Gramian matrices considered in Control Theory (see next subsection).

Instead of using the matrix $\mathcal W$ as a weighting, we may alternatively employ it akin to a `null model' term.
For instance, we can chose $\mathcal{W}$ to project out the average of $\mathbf{y}$ or certain other components (see also Appendix~\ref{sec:SI_relationshipsLouvain}).
If the system~\eqref{eq:gen_setup} corresponds to a diffusion processes, by chosing an appropriate projection, we can recover concepts such as the modularity matrix and its generalisations as specific cases (see \Cref{sec:SI_relationshipsLouvain,sec:SI_relationsDiffusion}).
However, the above formulation can equally be applied for other dynamics, as we will showcase in \Cref{sec:dim_reduction_ranking,sec:SItribes,sec:func_modules}.

\subsection{Interpretations of dynamical similarites and distances}
Before discussing specific applications of the above dynamical similarity measures, let us examine some properties of the above measures in more detail.

First, note that the definition of the dynamical similarity \Cref{eq:Psi} can be written in the form:
\begin{equation}
    \Psi(t) =  \mathcal B^\top \Xi(t) \mathcal B,
\end{equation}
where $\Xi(t)$ is governed by the following Lyapunov matrix differential equation~\cite{Abou-Kandil2012}:
\begin{flalign}
\label{eq:lyapunov_xi}
\frac{d\;\Xi}{dt} &= \mathcal{A}^\top\Xi +\Xi\mathcal{A}, \quad \text{with} \quad \Xi(0) = \mathcal{C}^\top\mathcal{WC}.
\end{flalign} 
Thus, $\Psi(t)$ is a dynamically evolving positive semi-definite Gram matrix, or a dynamic kernel matrix.
The same type of Lyapunov equation also governs the evolution of the covariance matrix of the system~\eqref{eq:gen_setup} driven by white Gaussian noise~\cite{Abou-Kandil2012,Skelton1997}, which yields another interpretation of the above similarity measure.

\subsubsection{Integrated dynamical similarity and control-theoretic interpretations of \texorpdfstring{$\Psi(t)$}.}
Instead of selecting a particular time $t$ in our similarity measure, we may integrate over time and thus define the integrated similarity measure
\begin{equation}\label{eq:Psi_int}
    \Psi_{[0,t]} := \int_0^t \Psi(t) dt.
\end{equation}
Analogously, we define the associated integrated squared distance matrix:
\begin{equation}\label{eq:D_int}
    D_{[0,t]}^{(2)} = \mathbf{1}\mathbf{z}^\top + \mathbf{z}\mathbf{1}^\top - 2\Psi_{[0,t]},
\end{equation}
where $\mathbf{z} = \text{diag}({\Psi_{[0,t]}})$ is the column vector containing the diagonal entries of $\Psi_{[0,t]}$ (cf.~\Cref{eq:distance}).
Only longer lived features will contribute significantly to this integral, and thus short-lived features are integrated out.
If we are mostly interested in features that are dominant over a certain range of time-scales we may thus employ the integrated similarity.

Note that for a diffusion dynamics on an undirected graph with $\mathcal A = -L, \mathcal C = I- \mathbf{11}^\top/n$, the distance measure $D_{[0,t]}^{(2)}$ is simply proportional to the resistance distance $\kappa_{ij}$ between nodes $i$ and $j$~\cite{Fouss}, i.e., 
\begin{equation}
    \lim_{t\rightarrow \infty} {\left [D_{[0,t]}^{(2)} \right]}_{ij} = \frac{\kappa_{ij}}{2} = \frac{1}{2}{(\mathbf{e_i}- \mathbf{e_j})}^\top L^\dagger (\mathbf{e_i}- \mathbf{e_j}),
\end{equation} 
where $L^\dagger$ is the Moore-Penrose pseudoinverse of the Laplacian, and $\mathbf{e_i}$ is the $i$-th unit vector.

The integrated Gramian $\Psi_{[0,t]}$ in~\eqref{eq:Psi_int} can also be interpreted in terms of an observability / controllability Gramian considered in Control Theory.
Specifically, consider the Gram matrix based on the $\mathcal L_2$ inner product:
\begin{equation*}
    \langle \mathbf{f_i}, \mathbf{f_j} \rangle_{\mathcal{L}_2} = \int_0^t \mathbf{f^\top_i} \mathbf{f_j} \, dt
\end{equation*}
between the vector functions $\mathbf{f_i}: [0,t] \rightarrow \mathbb{R}^n$ defined via the mapping $\mathbf{f_i}:~t\mapsto \mathcal{C} \, e^{\mathcal{A} t} \, \mathbf{e_i}$.
This is precisely the finite-time observability Gramian $G_O(t)$ of the linear system~\eqref{eq:gen_setup}, which is defined as:
\begin{equation}
\label{eq:obs_gramian}
G_O(t) = \int_0^t e^{\mathcal{A}^\top t} \, \mathcal{C}^\top\mathcal{C} \, e^{\mathcal{A} t} \, dt.
\end{equation}
${[G_O]}_{ij}$ quantifies how inferable the initial state at node $i$ is from output $j$.
Hence, a high value of the entry ${[G_O]}_{ij}$ signifies that node $j$ is highly observable from node $i$, when $C = I$.
More precisely, each entry reflects  how the energy of the initial states (localized on the nodes) spread to the outputs~\cite{Verriest2008}.
From our discussion above we may alternatively say that the observability Gramian~\eqref{eq:obs_gramian} measures the similarity between two nodes in terms of their dynamical response over the interval $[0,t]$.

Accordingly, $\Psi(t)$ can be interpreted as an instantaneous Gramian corresponding to a particular time instance $t$, i.e., $\Psi(t)$ may be understood as computing inner products between \emph{sampled} zero-state impulse response trajectories $t\mapsto \mathcal{C} \, e^{\mathcal{A} t} \mathcal B\mathbf{e_i}$ at a particular time $t$.
Indeed, our measure $\Psi(t)$ can be rewritten as
\begin{equation*}
    \Psi(t)= \mathcal B^\top \frac{d G_O(t)}{dt} \mathcal B.
\end{equation*}
As $G_O$ has the interpretation of an energy, the entries of $\Psi(t)$ may thus be interpreted as a power transferred between the nodes.

It is well known that there exists a duality between the observability of a system and the controllability of the system governed by transposed matrices.  
We may thus also view $\Psi(t)$ as assessing the instantaneous controllability of a dual system to~\eqref{eq:gen_setup} obtained by making the transformation $(\mathcal A, \mathcal B, \mathcal C) \rightarrow (\mathcal A^\top, \mathcal C^\top, \mathcal B^\top)$.
In a similar vein, we can explore the dual controllability measure in that system. 
A more detailed investigation of these directions will be the object of future work.

\subsubsection{Relations to time-scale separation, low-rank structure and model reduction. }
Asymptotically, the dynamics of many networked systems converges to a lower dimensional manifold.  
Think, for instance, of synchronisation processes.
In structured networks, however, one typically observes that the state transition matrix, and therefore the similarity matrix $\Psi (t)$, becomes numerically low-rank at much early times. 
Stated differently, in many structured networks we observe time scale separation linked to low dimensional subspaces of slowly decaying metastable states.
Therefore, the system can be effectively described by a small set of slow modes that govern the dynamics over some time scale. 
A feature specific to networked systems is the fact that these slow modes can be \textit{localized} on the space of nodes. 
It then follows, that instead of having to account for the whole system, we may just keep track of a few aggregated `metanodes', whose state is governed by the slow modes, thereby reducing the complexity of the dynamics. 

For a Laplacian diffusion dynamics ($\mathcal{A}=-L$) this idea can be made more precise using so-called externally equitable partitions~\cite{OClery2013}, which explicitly relate our similarity measure to model reduction.
Consider an external equitable partition (EEP) characterized by the relation
\begin{equation}
    L  H_{\text{EE}} = H_{\text{EE}} \widehat{L},
\end{equation}
where $H_{\text{EE}}$ is an indicator matrix encoding the EEP, and 
\begin{align}
\label{eq:quotient_Laplacian}
\widehat L = {(H_{\text{EE}}^\top H_{\text{EE}})}^{-1} H_{\text{EE}}^\top L H_{\text{EE}} = H_{\text{EE}}^+ L H_{\text{EE}}
\end{align}
is the Laplacian of the quotient graph, the graph in which each group of the partition becomes a `metanode'. 

It can be shown that if we observe such a system through its projection onto this external equitable partition (i.e., we set $\mathcal C= H^+_\text{EE}$), then every node within a group will have exactly the same influence on the observed output trajectories.
The similarity matrix $\Psi(t)$ can be written in terms of the quotient graph as:
\begin{align*}
    \Psi(t)  &=  {\exp(-Lt)}^\top {(H_\text{EE}^+)}^\top  H_\text{EE}^+ \exp(-L t) \\
             &={\left[\exp(-\widehat{L}t)  H_{\text{EE}}^+ \right]}^\top \exp(-\widehat{L}t)  H_{\text{EE}}^+,
\end{align*}
which shows that $\Psi$ will be \emph{block-structured}. 
Consequently the dynamics of the full system within the subspace spanned by the partition can be described \textit{exactly} by a reduced model~\cite{Monshizadeh2014, OClery2013}, which is governed here by $\mathcal A = \widetilde{L}, \mathcal C = I$ and has only a single input per group, equal to the average input within the original group.

It is instructive to compare the above dynamical block-structure to the notions like stochastic block-models~\cite{Holland1983,Snijders1997}, in which each node in a group has statistically the same (static) connection profile.
Here we are interested in nodes that have \emph{dynamically} the same effect, and define nodes accordingly. 
Note however, that the connections formed by each node do not have to be the same, but simply lead to similar dynamical effects: our measures assess the node similarity with respect to some observable $\mathbf{y}$ and not with respect to the connections formed.
In other words our objective is to obtain a joint low-dimensional description of the dynamics of the system and localized features of the network structure.
For the same network we may have different types of dynamical modules, depending on the dynamics acting on top of it.
In section~\Cref{sec:func_modules}, we will see  an example in which the structural grouping and the dynamical grouping of the nodes are indeed different.

\section{Dimensionality reduction using dynamical distances}\label{sec:dim_reduction_ranking}

\begin{figure*}[tb!]
 \centering
 \includegraphics{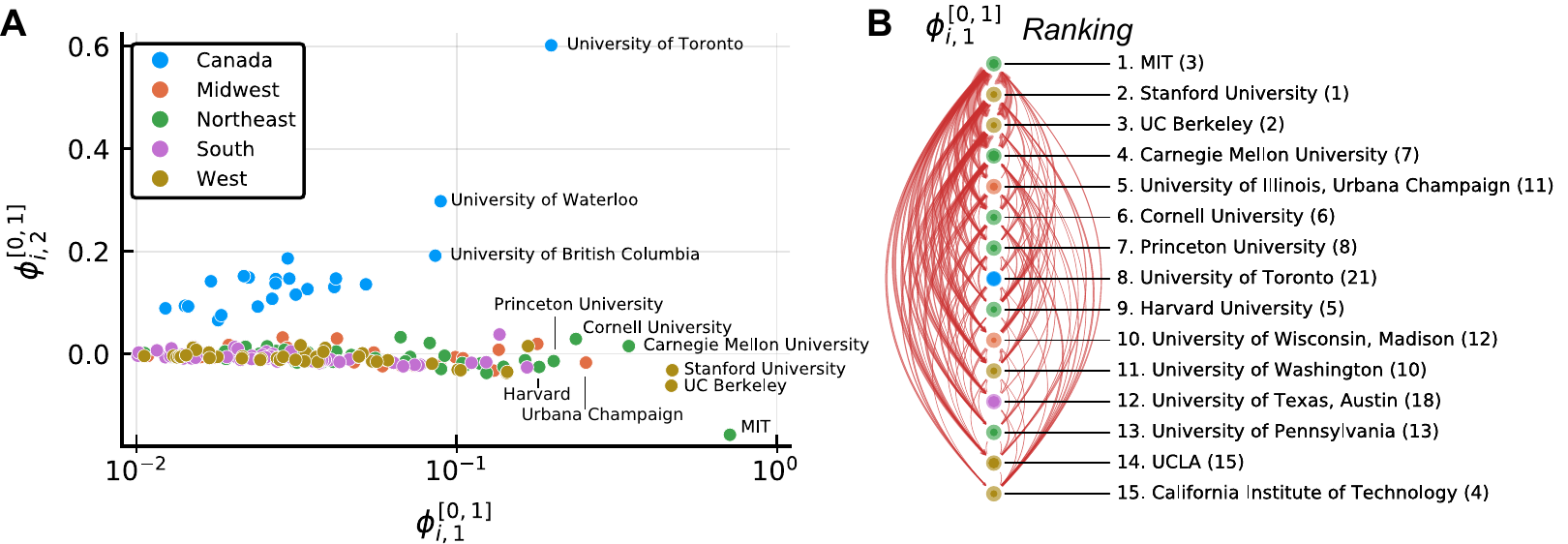}
 \caption{\textbf{Analysing academic influence using low-dimensional embeddings} 
     \textbf{A} We develop a low-dimensional embeddings based on the influence dynamics in the hiring network, as described in the text. The first two dimensions of this embedding are plotted. The first coordinate $\bm\phi_{i,1}^{[0,1]}$ is strongly correlated with the prestige ranking of Clauset et al.~\cite{Clauset2015}, highlighting the influential role played by the universities at the top. Interestingly, the second dimension distinguishes the Canadian universities from the US universities, showing that although these universities are well integrated within the faculty hiring market~\cite{Clauset2015}, they play a different role and exert a different type of influence on the system.
     \textbf{B} The ranking obtained when projecting onto the first coordinate only and the associated subgraph of faculty hirings. 
     The numbers in parenthesis correspond to the rankings obtained by Clauset et al~\cite{Clauset2015}.
     The arrows are proportional to the number of faculty moving between the institutions.
     Arrows pointing downwards in the ranking are plotted on the left, arrows point upward in terms of the ranking are plotted on the right. 
     The number of hirings from inside the institution itself (self-loops) are indicated by size of the core (darker color) of each node. 
     The area of the core is proportional to the number of self-loops compared to the total out-degree within this subnetwork.
 }%
 \label{fig:fac_hier}
\end{figure*}

In this section we outline how the above dynamical similarity measures may be employed for dimensionality reduction. 

Consider the spectral decomposition of $\Psi(t)$ into its eigenvectors $\mathbf{v}_{1}(t), \mathbf{v}_{2}(t), \ldots, \mathbf{v}_{n}(t)$ with associated eigenvalues 
$\mu_1(t) \geq \mu_2(t) \geq \cdots \geq \mu_n(t)$.
We define the mapping  $i \mapsto \bm{\phi}_{i}(t)$:
\begin{align}
\label{eq:embedding}
\bm{\phi}_{i}(t) 
=  {[\sqrt{\mu_1} \, v_{1,i},  \sqrt{\mu_2} \, v_{2,i}, \ldots, \sqrt{\mu_n} \, v_{n,i}]}^\top.
\end{align}
Using simple algebraic manipulations, it can now be shown that our dynamical distance measure~\eqref{eq:distance} can be written as:
\begin{equation}
    D_{ij}^{(2)}(t) = \|\bm{\phi}_{i} (t) -\bm{\phi}_{j}(t)\|^2.
\end{equation}
Hence, the vectors $\bm{\phi}_{i}$ map the data into a Euclidean space, in which the (Euclidean) distance is aligned with the dynamical impacts of the nodes at time $t$.
The entry-wise squared distance matrix $D^{(2)}(t)$ can thus be approximated by keeping only the first $c$ coordinates in each mapping $\bm{\phi}_{i}(t)$, thereby producing a low dimensional embedding of the original system.

For a diffusion dynamics with either $-\mathcal{A} = L$ (the combinatorial Laplacian) or $-\mathcal{A}^\top = L_\text{rw}$ (the random walk Laplacian matrix), it can be shown that $D^{(2)}(t)$ corresponds precisely to the distance induced by diffusion maps if the weighting matrix $\mathcal W$ is chosen appropriately~\cite{Coifman2005,Lafon2006}.
To see this, note that from the orthogonal spectral decomposition $L= \sum_i \lambda_i \mathbf{v}_i \mathbf{v}_i^\top$, it follows that 
\begin{equation*}
    \bm{\phi}_{i}(t) = {[e^{-\lambda_1 t} v_{1,i}, \ldots,e^{-\lambda_n t} v_{n,i}]}^\top 
\end{equation*}
are a time-dependent diffusion map embedding~\cite{Coifman2005,Lafon2006}.

\subsection*{Analysing academic hiring networks via low-dimensional embeddings}\label{sec:SIfaculty}
Which universities are the most prestigious in North America? 
In a recent study, Clauset et al.~\cite{Clauset2015} provided a data-driven assessment of this question by examining the hiring patterns of US-universities by means of a minimum violation ranking.
This ranking aims to order universities such that the fewest number of directed links, corresponding to faculty hirings, move from lower-ranked to higher-ranked universities.
Stated differently, universities with a higher prestige are assumed to act as sources of faculty for lower-ranked universities.

The dataset was released by Clauset et al.\footnote{\url{http://tuvalu.santafe.edu/~aaronc/facultyhiring/}} and consists of the placement of nearly 19,000 tenure track or tenured faculty among 461 North American departmental or school level academic units. 
The hiring data was collected for the disciplines business (112 institutions), history (144 institutions), and computer science (205 institutions).
In contrast to the history and business data, the computer science hiring data included 23 Canadian institutions.

Here we reconsider the ranking question using our above defined distance measure.
To illustrate our procedure, let us focus on the computer science (CS) data first.
Consider the adjacency matrix $A$ of the graph of hiring patterns for CS, where $A_{ij}$ denotes the number of faculty moving from university $i$ to university $j$.
This provides us with a directed, weighted network with 205 nodes corresponding to CS units at the departmental or school level, where we ignore movements of faculty to/from entities outside this set of 205 units.

Let us denote the influence of university by the state variable $x_i$.
We posit that a university $i$ exerts an influence on another university $j$ by sending faculty members to it.
To normalize for the size of the universities we divide this influence by the in-degree of each university, i.e., an influence of size 1 may be exerted on each university.
This leads us to consider an influence dynamics of the form 
\begin{equation*}
    {\dot{\mathbf{x}} = [K_\text{in}^{-1}A^T - I] \mathbf{x}}
\end{equation*}
among the universities, where ${K_\text{in} = \text{diag}(A^\top \mathbf{1})}$ is the diagonal matrix of in-degrees.
Note that one could also consider alternative artificial dynamics here, e.g., the relaxation dynamics of the recently proposed `spring rank' formalism, whose long-term behavior would then correspond to the spring rank~\cite{DeBacco2017}.

\begin{figure*}[tb!]
 \centering
 \includegraphics{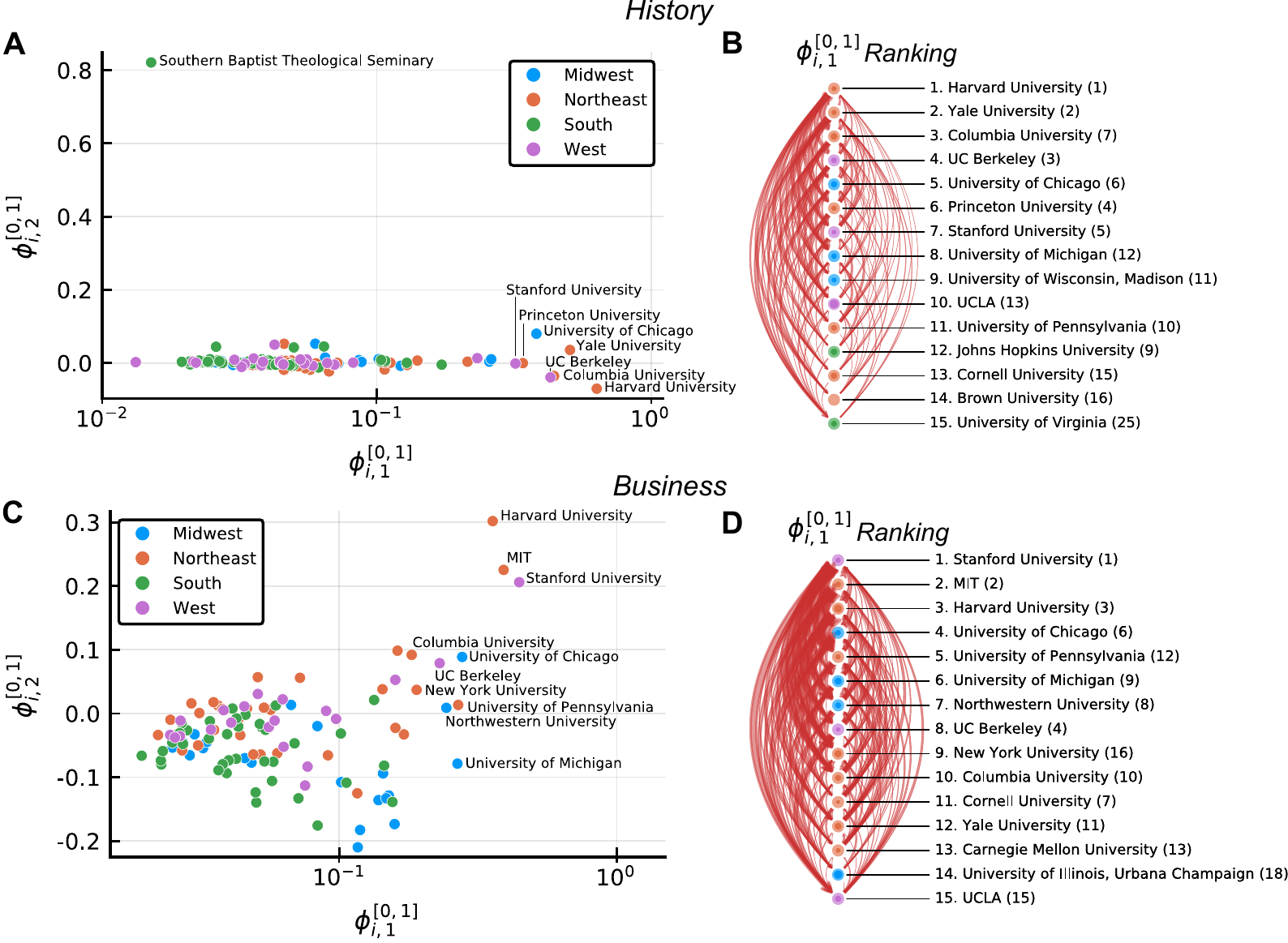}
 \caption{\textbf{Analysing academic influence using low-dimensional embeddings} 
     \textbf{A} Low-dimensional embeddings based on the influence dynamics in the hiring network, as described in the text (see Figure~\ref{fig:fac_hier}) for the discipline History.
     \textbf{B} The ranking obtained when projecting onto the first coordinate only and the associated subgraph of faculty hirings (see Figure~\ref{fig:fac_hier}). The Spearman rank correlation to the results obtained by Clauset et al is $\rho \approx 0.92$.
     \textbf{C} Low-dimensional embeddings based on the influence dynamics in the hiring network, as described in the text (see Figure~\ref{fig:fac_hier}) for the discipline Business.
     \textbf{D} The ranking obtained when projecting onto the first coordinate only and the associated subgraph of faculty hirings (see Figure~\ref{fig:fac_hier}). The Spearman rank correlation to the results obtained by Clauset et al is $\rho \approx 0.96$.
 }%
 \label{fig:fac_hier_SI}
\end{figure*}

As the hiring graph is not strongly connected, the long-term behavior will be dominated by a few modes depending on the initial condition.
We thus concentrate here on short time-scales for which paths of shorter lengths will be more important.
To avoid having to choose a particular time parameter, we integrate with respect to $t\in [0,1]$.
Note that while the underlying network is not strongly connected, there is no need to introduce a teleportation into the dynamics as is commonly the case in diffusion based methods.

To derive a low-dimensional embedding we approximate the resulting (squared) dynamical distance matrix $D^{(2)}_{[0,1]}$ via a low-rank spectral decomposition of $\Psi_{[0,1]} = V\Lambda V^T$.
To this end we define $\bm{\phi}_i^{[0,1]}$ via the relation 
\begin{equation}
    [\bm{\phi}_1^{[0,1]},\ldots,\bm{\phi}^{[0,1]}_n] = \Lambda ^{1/2}V^T =: \Phi_{[0,1]}.
\end{equation}
The vectors $\bm \phi_i^{[0,1]}$ define a new coordinate system, whose coordinates are ranked according to their importance to the dynamics.
We note that 
\begin{equation}
    {\left[D_{[0,1]}^{(2)} \right]}_{ij} = \left \|\bm{\phi}_i^{[0,1]} - \bm{\phi}_j^{[0,1]} \right\|^2, 
\end{equation}
and thus our dynamical distance can be approximated by truncating our coordinate system to the first few components of the vectors $\bm \phi_i(t)$.

Figure~\ref{fig:fac_hier} shows the results of this procedure when applied to the CS dataset of Clauset et al.~\cite{Clauset2015}.
We find that the first coordinates $\bm\phi_{i,1}^{[0,1]}$ are strongly correlated with the previously obtained ranking~\cite{Clauset2015} (Spearman rank-correlation $\rho \approx 0.90$), i.e., our dimensionality reduction maintains the essential features of the identified prestige hierarchy.
In addition, our embedding reveals that the Canadian universities play a somewhat different role in the system. 
Indeed the second coordinate $\bm{\phi}_{i,2}^{[0,1]}$ is singling out Canadian universities, highlighting that not all features of the influence dynamics are captured well by a unidimensional ranking (see Figure~\ref{fig:fac_hier}).
When symmetrizing the network, the Spearman correlation of the first dimension with the minimum violation ranking of Clauset et al.~\cite{Clauset2015} drops markedly to $\rho \approx 0.80$, emphasizing again that the directionality in this network is an essential feature.

In \Cref{fig:fac_hier_SI} we show the corresponding analyses for the disciplines history and business.
As shown, from the first dimension of the embedding we can again derive an influence ranking that is strongly correlated to the results obtained by Clauset et al. 

With Canadian institutions absent from the data for History and Business, the second dimension of the embedding appears to not correlate clearly with a geographical feature. 
For the History dataset, the Southern Baptist Theological Seminary is singled out in our second projection coordinate.
One of the main differences of this unit is its relatively large number of self-loops in the hiring data (11 hirings come from the same institution), leading to a highly localized influence of this institution.
Indeed, the coordinate of all other institutions is essentially zero in this second embedding dimension.

For the business data there is a slight separation along the second dimension.
More coastal regions (West, Northeast) tend to have a higher $\bm\phi_{i,2}^{[0,1]}$ projection.
South and Midwest institutions tend to have a lower coordinate $\bm\phi_{i,2}^{[0,1]}$.
However, the separation of the 3 top institutions may be better explained by their relative position in the network.
First, these are the only 3 institutions that placed more than 300 faculty members (outdegree 412 Stanford, 364 MIT, and 344 Harvard).
The next largest institution in terms of this placement is the University of Michigan (outdegree 282).
This large direct influence is further boosted, as not only are there strong ties from these top 3 institutions to most lower ranked universities, but also a relatively strong circular influence among these three top institutions.
A substantial fraction of the hirings of each of these 3 institutions comes from within their own small `rich club'.

While we focussed here on the first two embedding dimensions, there is no reason to restrict ourselves to 2 dimensional projections a priori.
Indeed the very same procedure can be applied to more dimensions, which could lead to a more nuanced appraisal of the relative influence of these institutions in the hiring network.
Our focus here was on the conceptual aspects of these embeddings, but a more detailed investigation, potentially linking these results to the relaxation dynamics of the recently proposed SpringRank method~\cite{DeBacco2017}, would be an interesting subject of future investigations.

\section{Dynamical embeddings of signed social interaction network}\label{sec:SItribes}
Many networked systems contain both attractive and repulsive interactions. Examples include social systems, in which people may be friends or foes, or genetic networks, in which inhibitory and excitatory interactions are commonplace.
Such systems can be represented as \textit{signed graphs}, with positive and negative edge weights.
A simple model for opinion formation on signed networks is given by~\cite{Altafini2013,Altafini2015}:
\begin{equation}\label{eq:signed_diffusion}
 \mathbf{\dot{x}} = -L_s \mathbf{x} + \mathbf{u},
\end{equation} 
where the \textit{signed Laplacian} matrix is defined as ${L_s = D_s -A_s}$ and the state vector $\mathbf{x}$ describes the `opinion' of each node.

Here $A_s$ is the adjacency matrix of the network, with positive and negative edge weights, and $D_s$ is the matrix containing the weighted absolute strengths of the nodes on the diagonal, ${[D_s]_{ii} = \sum_k |(A_s)_{ik}|}$ and $[D_s]_{ij}=0$ for  $i\neq j$. 
The signed Laplacian is positive semidefinite~\cite{Luca2010,Altafini2013} and reduces to the standard combinatorial Laplacian if $A_s$ contains only positive weights.
Clearly, this dynamics is of the form~\eqref{eq:gen_setup} discussed in the main text, with $\mathcal{A} = -L_s$ and 
$\mathcal{B}=\mathcal{C}=I$. 
In this case, the dynamic similarity
\begin{equation}
    \Psi (t) = {\exp(-L_s \, t)}^\top \exp(-L_s \, t),
\end{equation}
has time-independent eigenvectors $\mathbf{v}_i$ and associated eigenvalues $\mu_i(t)= e^{-\lambda_i t}$, where the 
$\mathbf{v}_i$ and $\lambda_i$ are eigenvectors and eigenvalues of $L_s$.

\begin{figure*}[tb!]
 \centering
 \includegraphics{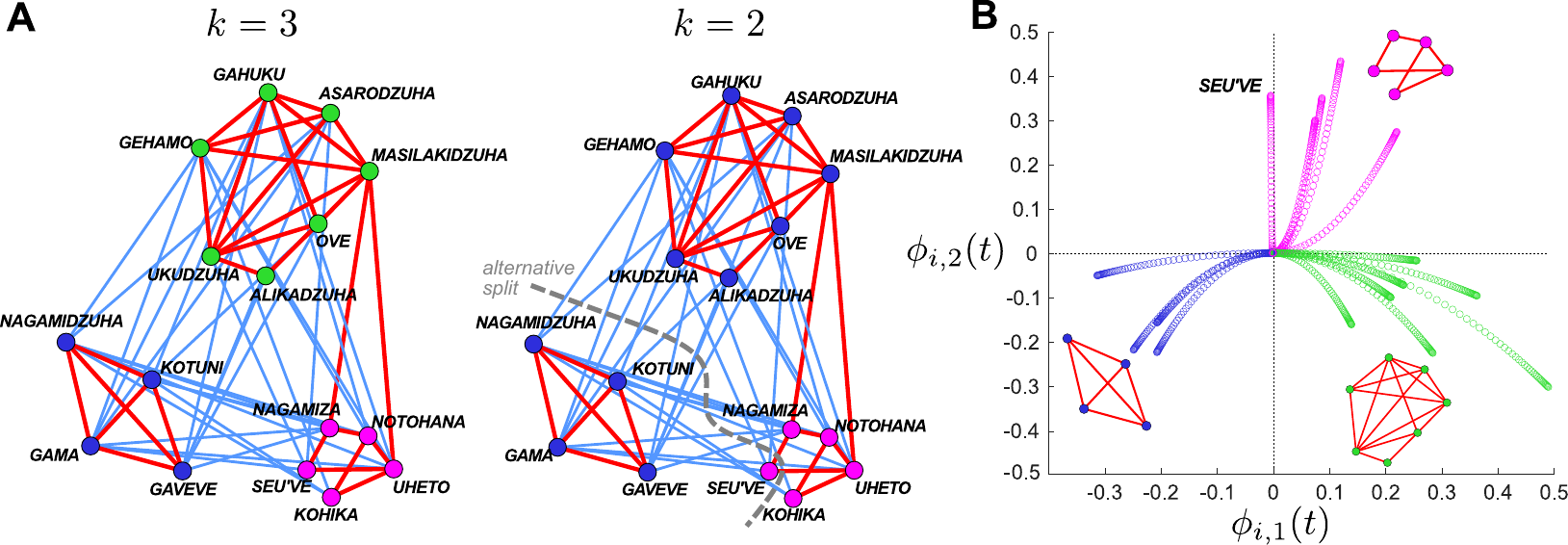}
 \caption{\textbf{Analysis of a signed social network: the highland tribes in New Guinea.} \textbf{A} The network of 16 tribes with positive interactions (`hina') in red and negative interactions (`rova') in blue. Spectral clustering using $c=2$ eigenvectors of the signed Laplacian $L_s$. The top 2 eigenvectors of $\Psi(t)$ reveals partitions into $k=3$ and $k=2$ groups with positive interactions mostly concentrated within groups, and antagonistic interactions across groups.  
     If we instead try to split the signed network into $k=2$ groups based on $c=1$ eigenvector, we obtain an alternative split as indicated by the dashed gray line.
 \textbf{B} The time evolution of the tribes in state space under the consensus dynamics~\eqref{eq:signed_diffusion} is represented through the dynamical embeddings $\bm{\phi}_i(t)$. Here we plot only the first two dominant coordinates. As time grows, the Seu've tribe switches from a marginal allegiance to the pink/green groupings (on the upper/lower right side in {B} with $\phi_{i,1}>0$) to be grouped with the blue block (left side in B with $\phi_{i,1}<0$). This is the result of an `enemy of my enemy is my friend' effect.}\label{fig:gama}
\end{figure*}

Let us consider the network of relationships between 16 tribal groups in New Guinea chartered by Read~\cite{Read1954} and first examined in the social network literature by Hage and Harari~\cite{Hage1983}. 
The relationships between the different tribes are either sympathetic (`hina'; red edges in Fig.~\ref{fig:gama}) or antagonistic (`rova'; blue edges in Fig.~\ref{fig:gama}). 
A `hina' edge signifies political alignment and limited feuds.
A `rova' edges denote relationships in which warfare is commonplace.

\subsection*{Spectral partitioning and dynamical embeddings}
To illustrate how our dynamical embedding can provide further insight into such a system with signed interactions, let us initially focus on a discrete categorization of the nodes into clusters, instead of finding a continuous embedding for our system.
Many methods have been proposed to cluster signed networks~\cite{Luca2010,Traag2009, Mucha2010} that can be used to find the groupings in the here considered setting.
All of them follow a combinatorial approach and aim to find dense groupings in the network containing a maximum number of positive links within the groups and most negative links across groups.
Perhaps the most straightforward way to split the nodes of the network into blocks is an approach based on spectral clustering~\cite{VonLuxburg2007}.
For signed networks, such a spectral clustering based on the signed Laplacian may be interpreted as optimizing a signed ratio cut~\cite{Kunegis2010}, which provides a principled way to detect groups in a signed network.

To split a system into $k$ groups, we assemble the matrix $V_c$ containing the $c$ eigenvectors corresponding to the smallest eigenvalues of $L_s$.
(Note that $V_c$ also corresponds to the $c$ dominant eigenvectors of $\Psi(t)$ for $t>0$.)
The rows of $V_c$ are then taken as new $c$-dimensional coordinate vectors for each node on which a $k$-means clustering is run to obtain the $k$ modules.  
Though, in general, the dimension of the coordinate space $c$ and the number of modules $k$ need not be the same, one typically chooses $c=k$ or $c=k-1$~\cite{VonLuxburg2007}.   
To showcase the utility of this procedure, we applied this form based on the 2 dominant eigenvectors ($\mathbf{v}_1$ and $\mathbf{v}_2$) to split the network into $k=2, 3$ groups (Fig.~\ref{fig:gama}A).  
The blocks obtained are characterized by high internal density of positive links with negative links placed across groups.
Interestingly, if we aim to cluster the network into $k=2$ groups using only $c=1$ eigenmodes of $L_s$, we obtain a grouping in which the Seu've tribe is place together with the Gama, Kotuni, Gaveve, and Nagamidzuha tribe, and the remaining tribes form a second group (see Figure~\ref{fig:gama}).

To gain additional insight, we study the dynamical coordinates $\bm{\phi}_{i}(t)$ defined in Eq.~\eqref{eq:embedding}, which can be seen as feature vectors that combine the information of the eigenvectors and eigenvalues. 
The time evolution of these feature vectors provides a dynamical embedding of the signed opinion network, reflecting the relative position of the nodes (tribes) in the state-space of `opinions'.
Instead of providing a discrete categorization, the continuous nature of the embedding provides us with a more nuanced view on how closely aligned individual tribes are to each other over time (Figure~\ref{fig:gama}A).
Note that the spectral clustering with $c=1$ discussed above corresponds essentially to the long-term behaviour of this dynamics.
Our dynamical embedding shows that the Seu've tribe has effectively a zero, but slightly negative coordinate within direction $\phi_{i,1}$.
Hence, if we concentrate only on $c=1$ eigenvector the obtained split will be commensurate with the $\phi_{i,1}$ coordinate, which is exactly the partition obtained before.

To understand why the impact of the Seu've tribe on the network in terms of the $\phi_{i,1}$ coordinate is indeed negative, it is instructive to examine the position of Seu've in the network in a bit more detail.
Note that the Seu've tribe has 2 direct positive links with the Nagamiza and the Uheto tribe.
Seu've has also negative with the Ukurudzuha, Asarodzuha and the Gama tribe.
The split into 3 groups is exactly aligned with these positive and negative relationships.
The spectral split into 2 groups based on the first dominant vector appears to be at odds with these relationships, though.

The reason for this at first sight non-intuitive split is a behavior of the type ``the enemy of my enemy is my friend'', which is inherent to signed interaction dynamics. 
This effect plays a more important role for larger time scales and is thus reflected in the sign patters of the dominant eigenvector.
In this case, the mutual antipathy of all three Seu've, Gama and Nagadmidzuha clans against the Asarodzuha tribe implies that Gama, Nagadmidzuha and Seu've behave in the long run similar and thus have a negative $\phi_{i,1}$ coordinate.

Following structural balance theory~\cite{Cartwright1956}, one may conjecture that the Gama-Seu've relationship could cease to be of `rova' type in a future observation of the network.
In his socio-ethnographic characterization of this tribal system, Read indeed remarked that the system was ``relative and dynamic''~\cite{Read1954}.
Our analysis highlights that there is additional information to be gained when adopting a dynamical point of view, as shown by potential of the Seu've tribe to be `turned around'.

Indeed, instead of using the eigenvector of $L_s$, we may alternatively use the dynamical $\phi_{i}$ coordinates for clustering, thereby taking into account the eigenvalues of the dynamics as well.
For $k=3$ groups the resulting clustering is the same as the one we obtain from spectral clustering based on the eigenvectors of $L_s$ alone.
However, for $k=2$ the split is somewhat different.
For all but the largest time-scales the green and pink groups are merged, and only for very large time-scales does the Seu've tribe `flip' and become part of the group containing the Gama tribe.

\begin{figure*}[tb!]
 \centering
 \includegraphics{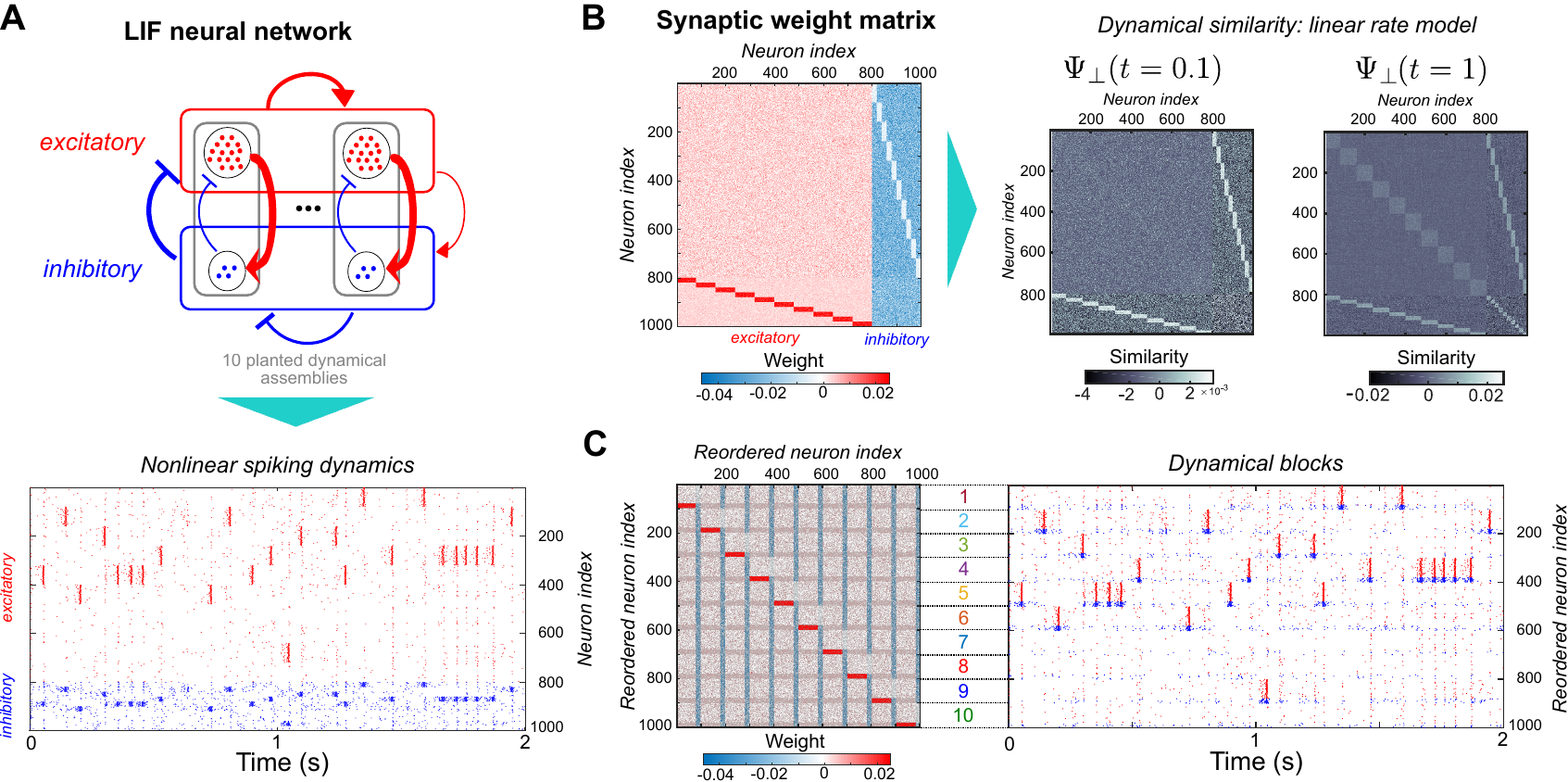}
 \caption{\textbf{Finding dynamical groups in a neural network description.} 
     \textbf{A} Schematic of the connectivity of the leaky integrate and fire neuronal network, which shows its disassortative feedback structure between inhibitory and excitatory neurons. The exemplar raster plot illustrating its spiking dynamics shows that the system is characterized by slow switching between coherent spiking activity of $10$ groups of neurons (each containing both inhibitory and excitatory units).
     \textbf{B} Left: The weighted, signed and directed synaptic connectivity matrix ($W_N$) of the network does not contain groups of nodes with high internal connection density. However, the analysis of the linear rate model governed by this connectivity matrix~\eqref{eq:rate_model} using the dynamical similarity~\eqref{eq:psi_centered} in conjunction with a Louvain-type optimization reveals the presence of 10 dynamical modules. Right: Visualization of the centered similarity~\eqref{eq:psi_centered}. For visualization purposes only the diagonal of $\Psi_\perp$ has been removed. Note that how after an initial short transient period the block structure into 10 groups becomes apparent, in comparison to the original weight matrix.
     \textbf{C}~The blocks revealed from the linear rate model coincide with the dynamically co-activated groups of neurons in the full LIF dynamics, as shown by reordering the neuron indices. On the original weight matrix, they correspond however to a mixture of the blocks inside the weight-matrix $W_N$.
 }\label{fig:neuro}
\end{figure*}

\section{Finding functional modules in neuronal networks via dynamical similarity measures}\label{sec:func_modules}
As a final example for the utility of our embedding framework, we now consider the analysis of networks of spiking neurons.
Specifically we will consider the dynamics of a network of leaky-integrate-and-fire (LIF) neurons.
Due to their computational simplicity yet complex dynamics, networks of LIF neurons are widely used as scalable prototypes of neural activity. 
Recently, it has been shown that LIF networks can display ``slow switching activity''~\cite{Litwin-Kumar2012, Schaub2015}, sustained in-group spiking that switches from group to group across the network. 
Importantly, the cell assemblies of coherently spiking neurons in this context can include both excitatory neurons and inhibitory neurons, and dense clusters of connections are not necessary to give rise to such dynamics (see Figure~\ref{fig:neuro}).
These cell assembles are thus an interesting example for a \emph{functional module}, that cannot be discerned from the network structure alone.

As has been shown previously, key insights into the nonlinear LIF dynamics can be obtained from linear rate models of the following form~\cite{Schaub2015}, which are amenable to the methodology developed above:
\begin{align}\label{eq:rate_model}
    \mathbf{\dot{x}} &= (- I + W_N) \mathbf{x} + \mathbf{u}, 
\end{align}
where $\mathbf{x}$ describes the $n$-dimensional firing rate vector relative to baseline; $\mathbf{u}$ is the input; and $W_N$ is the asymmetric synaptic connectivity matrix containing excitatory (positive) and inhibitory (negative) connections between the neurons. 
The asymmetry of $W_N$ follows from Dale's principle~\cite{Strata1999}, which states that each neuron acts either completely inhibitory or completely excitatory on its efferent neighbours.
Clearly, the rate dynamics~\eqref{eq:rate_model} is of the form~\eqref{eq:gen_setup}.

In this example we consider a LIF network, whose coupling matrix $W_N$ is shown by the signed network in Figure~\ref{fig:neuro}B.  
The structure of the network can be described by a block-partition into 20 blocks: 10 groups of excitatory neurons, and 10 groups of inhibitory neurons, whose ordering is consistent with with the network drawing in Figure~\ref{fig:neuro}B.
The connectivity patterns between these blocks are homogeneous in terms of the probability of observing a connection and their connection link-strengths.
If we were to partition this coupling matrix into homogeneously connected blocks in terms of weights and number of connections, we would find these 20 structural blocks. 

It turns out that this arrangement corresponds however to only 10 planted dynamical cell assemblies, each consisting of a mixture of inhibitory and excitatory neurons.
Thus, while from an inspection of $W_N$ we may conclude that there should be 20 groups, we know from our design that there only 10 dynamically relevant groupings~\cite{Schaub2015}.
In order to assess which dynamical role is played by the different neurons, we thus consider our dynamical similarity measure, this time however not with a focus on deriving an embedding, but with an eye towards identifying the planted functional groups in the (nonlinear) dynamics.

Since cell assemblies are characterized by a relative firing increase/decrease with respect to the population mean, we use a centered similarity matrix by chosing a weighting matrix of the form $\mathcal W = I - \mathbf{11}^\top/n$.
\begin{equation}
\label{eq:psi_centered}
    \Psi_\perp (t) = {\exp(\mathcal{A} t)}^\top \left( I - \frac{\mathbf{1}\mathbf{1}^\top}{n} \right) \exp(\mathcal{A} t),
\end{equation}
where $\mathcal A = (- I + W_N)$.
As discussed in \Cref{sec:SI_relationshipsLouvain}, this can be interpreted as a choice of a null model, or as introducing a relaxation on the distance matrix different to the low-rank approximation discussed in the previous section.
These type of relaxations of our dynamical similarity measures enable us to draw further connections to quality functions more commonly employed in network analysis, as we discuss in the next section.

\subsection*{Revealing dynamical modules with a Louvain-like combinatorial optimization}
Let us consider a general similarity matrix $\Psi$ defined via an orthogonal projection $\mathcal{W}_{\perp} =I - \bm{\nu}\bm{\nu}^\top$ as weighting matrix:
\begin{equation}\label{eq:Psi_perp}
    \Psi_\perp(t) =  \mathcal B^\top {\exp(\mathcal At)}^\top\mathcal{C}^\top \mathcal{W_\perp} \, \mathcal{C}\exp(\mathcal A t) \mathcal B.
\end{equation}
Note that the centered similarity~\eqref{eq:psi_centered} considered for our neuronal network is precisely of this form.

Clearly, the weighted inner product $\langle  \mathbf{y}_{i}(t), \mathbf{y}_{j}(t)  \rangle_{_\mathcal{W}}$ projects out particular properties associated with $\bm{\nu}$.
The choice of $\mathcal{W}_{\perp}$ can thus be interpreted as selecting a type of  `null model' for the nodes.
Alternatively, we can think of this operation as projecting out uninformative dimensions of the data, thereby providing a geometric perspective on the selection of a null-model (see \Cref{sec:SI_relationshipsLouvain,sec:SI_relationsDiffusion}).
For instance, choosing $\bm{\nu} = \mathbf{1}/\sqrt{n}$ as done in~\eqref{eq:psi_centered} is equivalent to centering the data by subtracting the mean of each of the vectors $\mathbf{y}_i(t)$.

Having defined a similarity matrix $\Psi_{\perp}(t)$ as above, we can obtain dynamical blocks with respect to the null model $\bm{\nu}$ as follows. 
Let us define the quality function
\begin{equation}
\label{eq:quality}
    r_{\bm{\nu}}(t,H) = \text{trace } H^\top \Psi_{\perp}(t) \, H,
\end{equation}
where $H$ is a partition indicator matrix with entries $H_{ij}=1$ if state $i$ is in group $j$ and $H_{ij}=0$ otherwise. 
The combinatorial optimization of  $r_{\bm{\nu}}(t,H)$ over the space of partitions can be performed efficiently for different values of the time $t$ through an augmented version of the Louvain heuristic.

We applied this optimization procedure for the quality function induced by the similarity matrix~\eqref{eq:psi_centered} to search for possible functional modules within the neuronal network.
As can be seen in Figure~\ref{fig:neuro}, optimizing~\eqref{eq:quality} reveals precisely the mixed groups of excitatory and inhibitory neurons that exhibit synchronized firing in the fully non-linear LIF network simulations.
Again, these groups do not correspond to tightly knit groups in the topology (see Figure~\ref{fig:neuro}) but rather reflect dynamical similarity.

As our example highlights, if we are interested in some kind of process on a network, rather than the network structure itself, using a dynamical similarity measure can lead to a more meaningful analysis.
The specific example here is however not meant to suggest a particular null model, or a generic optimization method.
Indeed, similar results can be obtained, e.g., by directly analysing $\Psi$ (or $D^{(2)}$) using spectral techniques as outlined above.

\section{Discussion}\label{sec:discussion}
Building on ideas from systems and control theory, we have presented a framework that provides dynamical embeddings of complex networks, including signed, weighted and directed networks. 
These embeddings can be used in a variety of analysis tasks for network data.
We have focused here on applications to dimensionality reduction and the detection of dynamical modules to highlight important features of our embedding framework.
However, the dynamical similarity measures $\Psi(t)$ and $D^{(2)}(t)$ may also be used in the context of other problem formulations not considered here.
For instance, we could consider the (functional) networks induced by our dynamical similarity measures, and employ generative models~\cite{Holland1983,Newman2015,Peixoto2013} for their analysis.
One way to approach this would be to define a (negative) Hamiltonian based on our similarity matrix, e.g., in a form similar to~\eqref{eq:quality}, and a Boltzmann distribution of the corresponding form.
In this view the state-variables of the node (or other labels defined on the nodes) would correspond to latent variables that are coupled via the Hamiltonian.

One may further consider the extension to kernels computed directly from nonlinear dynamical systems, akin to the perturbation modularity recently introduced by Kolchinsky et al.~\cite{Kolchinsky2015}, or consider linearisations around a particular state of interest. 
Alternatively, $\Psi(t)$ could be extended to represent nonlinear systems through an inner product in a higher dimensional space, e.g, by using the `kernel trick'~\cite{Schoelkopf2002}.
Our measures also provides links with other notions of similarity in networks including structural-equivalence, diffusion-based~\cite{Kondor2002,Smola2003} and iterative node similarity in networks~\cite{Blondel2004,Leicht2006}.
Such connections are interesting for machine learning, where a good measure of similarity is central to solving problems such as link prediction~\cite{Lue2011} and node classification~\cite{Fouss}.

For simplicity, we assumed in the examples above the number of state variables equals the number of nodes in the network.
Nevertheless, our derivations remain valid when there is more than one state variable per node. 
For instance, our ideas may be readily translated to multiplex networks~\cite{Mucha2010} or networks with temporal memory~\cite{Rosvall2014,Delvenne2015}, that feature expanded state space descriptions and have gained considerable interest recently.

We remark that the measures presented here are different from correlation analysis of time-series data, as considered, e.g., by~\citet{MacMahon2015}.
Instead of interpreting a correlation matrix as a functional network from $n$ scalar valued time-series and then analysing this correlation matrix, we start with the joint description of a network and a dynamics.

Conceptually, the similarity measure $\Psi(t)$ has strong theoretical links to model reduction and controllability, which provide meaningful interpretations of dynamic blocks in terms of coarse-grained representations. 
Classic model reduction~\cite{Dullerud2000,Baur2014,Schilders2008} aims to find reduced models that approximate the input-output behavior of the system;
yet the states of the reduced model do not usually have a sparse support in terms of the states of the original system. 
In contrast, the dynamical blocks found using $\Psi$ are directly associated with particular sets of nodes and can thus be \textit{localized} on the original graph, an important requirement for many applications.  
Future work will investigate alternative measures to $\Psi(t)$ based on the duality between controllability and observability Gramians from control, as well as measuring the quality of the dynamical blocks in a model reduction sense.

\acknowledgments{JCD, and RL acknowledge support from: FRS-FNRS\@; the Belgian Network DYSCO (Dynamical Systems, Control and Optimisation) funded by the Interuniversity Attraction Poles Programme initiated by the Belgian State Science Policy Office; and the ARC (Action de Recherche Concerte) on Mining and Optimization of Big Data Models funded by the Wallonia-Brussels Federation.
MTS received funding from the European Union’s Horizon 2020 research and innovation programme under the Marie Sklodowska-Curie grant agreement No 702410.
MB acknowledges funding from the EPSRC (EP/N014529/1).
The funders had no role in the design of this study; the results presented here reflect solely the authors' views.
We thank Leto Peel, Mauro Faccin, and Nima Dehmamy for interesting discussions.}
\bibliography{./references}

\clearpage
\appendix

\section{Leaky-integrate-and-fire neural networks with functional modules}\label{sec:SIneuro}
Due to their computational simplicity yet complex dynamics, networks of LIF neurons are widely used as scalable prototypes of neural activity. 
The non-linear dynamics of LIF models reproduce Poisson-like neuronal firing with refractory periods, among other features.
Here, we employ that LIF networks display structured behavior~\cite{Litwin-Kumar2012, Schaub2015}, in which sustained in-group spiking switches from group to group across the network. 
Importantly, these cell assemblies of coherently spiking neurons include both excitatory neurons (which exhibit a positive influence on their neighbours) and inhibitory neurons (whose influence is negative), which have different connection profiles.
Moreover, we remark that these groups are not densely connected clusters which are only weakly connected to other clusters, but the behavior emerges from the connections between the various groups.
Stated differently, the observed grouping is dynamical (functional) rather than structural.

We simulated leaky-integrate-and-fire (LIF) networks with $n=1000$ neurons ($800$ excitatory, $200$ inhibitory).
Using a time step of $0.1$ms, we numerically integrated the non-dimensionalized membrane potential of each neuron,
\begin{equation}\label{eq:LIF_model}
  \dfrac{d V_i(t)}{dt} = \dfrac{1}{\tau^{E/I}_m}(u_i - V_i(t)) + \sum_{j} {[W_{N}]}_{ij} \, g^{E/I}_j(t), 
\end{equation}
with a firing threshold of $1$ and a reset potential of $0$.
The input terms $u_i$ were chosen uniformly at random in the interval $[1.1, 1.2]$ for excitatory neurons, and in the interval $[1, 1.05]$ for inhibitory neurons. 
The membrane time constants for excitatory and inhibitory neurons were set to $\tau^E_m = 15$~ms and $\tau^I_m = 10$~ms, respectively, and the refractory period was fixed at $5$~ms for both excitatory and inhibitory neurons.
Note that although the constant input term is supra-threshold, balanced inputs guarantee an average sub-threshold membrane potential~\cite{Litwin-Kumar2012}.
The network dynamics is captured by the sum in~\eqref{eq:LIF_model}, which describes the input to neuron $i$ from all other neurons in the network  and ${[W_N]}_{ij}$ denotes the weight of the connection from neuron $j$ to neuron $i$.
Synaptic inputs are modelled by $g^{E/I}_j(t)$, which is increased step-wise instantaneously after a presynaptic spike of neuron $j$ ($g_j^{E/I}\rightarrow g_j^{E/I} +1$) and then decays exponentially according to:
\begin{equation}\label{eq:synapse}
    \tau_s^{E/I} \dfrac{d g_j^{E/I}}{dt} = - g_j^{E/I}(t),
\end{equation}
with time constants $\tau^E_s =3$ ms for an excitatory interaction, and $\tau^I_s =2$ ms if the presynaptic neuron is inhibitory.
Excitatory and inhibitory neurons were connected uniformly with probabilities $p_{EE}=0.2$, $p_{II}=0.5$, and weight parameters $W_{EE} = 0.022$ and $W_{II}= 0.042$, respectively.

The network comprised 10 functional groups of neurons, each of which consists of 80 excitatory and 20 inhibitory neurons connected as follows.
The excitatory neurons are statistically biased to target the inhibitory neurons in their own assembly with probability $p_{IE}^{in} = 0.90$ and weight $W_{IE}^{in} = 0.0263$, compared to $p_{IE} = 0.4545$ and $W_{IE} = 0.0087$ otherwise.
Inhibitory neurons connect to all excitatory neurons with probability $p_{EI}=0.5263$ and weight $W_{EI}= 0.045$, apart from the excitatory neurons in their own assembly which are connected with probability $p_{EI}^{in} = 0.2632$ and  $W_{EI}^{in} = 0.015$.  
Note that, while from a purely structural point of view we may split this network into 20 groups (10 groups of excitatory neurons, 10 groups of inhibitory neurons; see also Figure~\ref{fig:neuro}), it can be shown that this configuration gives rise to 10 functional groups of neurons firing in synchrony with respect to the rest of the network~\cite{Schaub2015}.

\section{Relations between dimensionality reduction and module detection}\label{sec:SI_relationshipsLouvain}
In this section we elaborate on the relationship between dimensionality reduction and the detection of dynamical modules as discussed in the main text.

Let us initially consider the problem from the point of view of the squared distance matrix $D^{(2)}$, where we omit writing the time-dependence to emphasize that the derivations below apply to both the integrated $D^{(2)}_{[0,t]}$ as well as the instantaneous distance matrix $D^{(2)}(t)$.

A naive idea to derive a clustering measure would be to simply try and place all nodes into the same group such that the sum of the distances in each group is minimized, which would lead to the following optimization procedure:
\begin{equation*}
    \min_H \text{trace } H^\top D^{(2)}H,
\end{equation*}
where $H\in{\{0,1\}}^{n\times k}$ is a partition indicator matrix with $H_{ij} = 1$ if node $i$ is in group $j$ and $H_{ij} = 0$ otherwise. 
We can rewrite the above using the definition of $D^{(2)}$ as 
\begin{equation*}
    \min_H \text{trace } H^\top\left[ \mathbf{1}\mathbf{z}^\top + \mathbf{z}\mathbf{1}^\top - 2\Psi \right] H,
\end{equation*}
where $\mathbf{z}=\text{diag}(\Psi)$ is the vector containing the diagonal entries of $\Psi$.
It is easy to see that if $k$ is not constrained in the above optimization problem, then the best choice will be to trivially put each node in its own group ($k=n$).
Stated differently, if we are free to choose any number of groups $k$, then we can make the distance within each group zero, thus minimizing the above objective.

One potential remedy to fix the above shortcoming would be to fix the number of groups, a priori, and then perform some kind of selection procedure afterwards to pick the number of groups.
Another option is to introduce some `slack' in the distance measurements, thus permitting nodes whose distance is comparably small to contribute negative to the cost function (which is here to be minimized).
As we will show in the following this naturally leads to a problem formulation akin to many network partitioning procedures which have been proposed in the literature.

Let us consider the spectral expansion ${\Psi = \sum_{i=1}^n \lambda_i \mathbf{v}_i\mathbf{v}_i^\top}$, where we assume the eigenvalues to be ordered, such that $\lambda_1 > \cdots > \lambda_n \geq 0$.
Then we can rewrite $D_{ij}^2$ as:
\begin{equation*}
    D_{ij}^2 = \sum_{k=1}^n \lambda_k {(\mathbf{v}_k^\top\mathbf{e_i})}^2 +\lambda_k {(\mathbf{v}_k^\top\mathbf{e_i})}^2 -2 \lambda_k (\mathbf{v}_k^\top\mathbf{e_j})(\mathbf{v}_k^\top\mathbf{e_i}),
\end{equation*}
where $\mathbf{e_i}$ is the i-th unit vector.
Let us now introduce some slack variables (multipliers) $\gamma_k$ for all the modes in the first two terms and rewrite the above expression in terms of the dynamical coordinates $\bm \phi_k$:
\begin{flalign*}
    D_{ij}^2 &= \sum_{k=1}^n \gamma_k\lambda_k {(\mathbf{v}_k^\top\mathbf{e_i})}^2 +\gamma_k\lambda_k {(\mathbf{v}_k^\top\mathbf{e_i})}^2 -2 \lambda_k (\mathbf{v}_k^\top\mathbf{e_j})(\mathbf{v}_k^\top\mathbf{e_i}) \\
             &= \sum_{k=1}^n \gamma_k[{(\phi_{i,k})}^2 +{(\phi_{j,k})}^2] -2 \bm \phi_{i}^\top\bm \phi_{j},
\end{flalign*}
which shows that $\gamma_k$, may be seen as weighting functions for the first $k$ coordinates in the $\bm \phi$ coordinate space for the first 2 (norm) terms in the distance. 

Using the above derivation, let us rewrite the previously considered minimization as an equivalent maximization problem.
\begin{flalign*}
    &\max_H \; 2 \; \text{trace } H^\top \left[\Psi - \frac{1}{2}(\mathbf{\tilde{z}}\mathbf{1}^\top - \mathbf{1}\mathbf{\tilde z}^\top)\right]H\\
    &  \text{with } \mathbf{\tilde{z}} = \text{diag } \Phi^\top \Gamma \Phi \in \mathbb{R}^n, \\
    & \qquad \Gamma = \text{diag}(\gamma_1, \ldots,\gamma_n) \in \mathbb{R}^{n\times n}
\end{flalign*}

While different weighting schemes $\{\gamma_k\}$ are of potential interest here, let us now consider the specific choice $\gamma_1 = 1$, $\gamma_k=0,  (k>1)$,
which corresponds to making a simple low rank-correction of $\Psi$.
This specific scheme is akin to choosing a type of null model in our optimization scheme as we will illustrate next.

For concreteness, let us consider the familiar case of Laplacian dynamics for a symmetric graph, i.e., $\Psi = {\exp(-Lt)}^\top\exp(-Lt)$.
In this case the eigenvectors of $\Psi$ are constant over time, and the first eigenvalue of $\Psi$ is one with an associated constant eigenvector and thus $\phi_{i,1} = 1/\sqrt{n}, \forall i$.
This leads to an optimization of the form:
\begin{flalign*}
    &\max_H 2 \; \text{trace } H^\top \left[\exp(-2Lt) - \frac{1}{2n}(\mathbf{1}\mathbf{1}^\top - \mathbf{1}\mathbf{1}^\top)\right]H,
\end{flalign*}
which can be simplified to the equivalent optimization:
\begin{flalign*}
    &\max_H \text{trace } H^\top \left [\exp(-2Lt) - \frac{1}{n}\mathbf{1}\mathbf{1}^\top \right]H
\end{flalign*}
Note that this is just the (rescaled) Markov stability at time $2t$ (see also Section~\ref{sec:SI_relationsDiffusion}).
Linearising the above expression thus leads to recovering a Potts-model like community detection scheme~\cite{Delvenne2013}, where the last term can be identified with an Erd\H{o}s-R\'enyi null model.
Following exactly the same procedure, similar expressions may also be derived for various other null models, such as the configuration model.

Further, as discussed in the next section, the above expression can be rewritten in the form $r_{\bm{\nu}}(t,H)$ (see Equation~\eqref{eq:quality}), and can thus be interpreted as a quality function that we can optimize using the Louvain optimization scheme.
This emphasizes how the Louvain optimization scheme can be interpreted as operating with an approximation of the distance matrix / similarity matrix $\Psi$.
This result may be used in several ways to derive some more general null models by choosing an appropriate weighting scheme.

\subsection{Null models, projections and time-parameter choices}
As observed above, the Louvain algorithm might be seen as solving a closely related (`dual') problem to the distance minimization.
However, instead of trying to find groups with minimal distance according to some criterion, the optimization operates in terms of the associated similarity measure (inner product).
An interesting question that we will not pursue in the following would thus be to investigate equivalent distance based optimization problems.
Instead, in the following we will discuss how the here derived formulation ties in with many quality function commonly considered in networks analysis.

Within network science many quality functions for community detection can effectively be written in the form~\cite{Fortunato2010,Reichardt2006,Delvenne2013,Campigotto2014}:
\begin{equation}
    r(H) = \text{trace } H^\top \left [G - \alpha N \right ]H,
\end{equation}
where $G$ is a term customarily related to the network structure, and $N$ is a `null-model' term, which customarily includes a scalar multiplier $\alpha$ as a free resolution parameter.
It is insightful to rewrite the above as:
\begin{equation}
    r(H) = \langle H, GH\rangle - \alpha \langle H, NH\rangle.
\end{equation}
In particular, if the null model term is a positive semi-definite matrix, we can further simplify this to:
\begin{equation}
    r(H) = \langle H, GH\rangle - \alpha \|N^{1/2}H\|_F^2,
\end{equation}
where $\|\cdot\|_F$ is the Frobenius norm.
This highlights how the null model term acts effectively as a regularization term (similar to regression type-problems) with a weighting factor (Lagrange multiplier) given by the resolution parameter $\alpha$.

Let us now consider how the above formulations apply to the similarity measures derived above.
To link with our above discussion let us concentrate on the case where we (partially) project out a rank-1 term via $\mathcal W = I - \alpha \bm{\nu}\bm{\nu}^\top$.
Note that the same effect can also be achieved by choosing $\mathcal C$ appropriately, providing additional interpretations which we will not explore in the following.
This would lead to a similarity matrix $\Psi_\perp$ of the form:
\begin{equation}
    \Psi_\perp = Y^\top Y - \alpha Y^\top\bm{\nu}\bm{\nu}^\top Y,
\end{equation}
where $Y=\mathcal{C}\exp(\mathcal{A} t)\mathcal B$.
This kind of similarity lead to a quality function of the form:
\begin{flalign}
    r_\nu(t,\alpha,H) &=  \|Y(t) H\|^2  - \alpha \|\bm{\nu}^\top Y(t) H\|^2\\
                      &= \|H\|_{\Psi(t)}^2 - \alpha \|\bm{\nu}^\top Y(t)H\|^2
\end{flalign}
where we explicitly have written out the dependency on $t$ and $\alpha$ and have defined the semi-norms $\|X\|_Y := \text{trace }X^\top Y X$, which is possible in our formulation as all the relevant matrices are positive semi-definite.

From the above we can make the following observations.
First, as already alluded to before, the influence of the resolution parameter $\alpha$ is akin to a Lagrange multiplier, which \emph{linearly} scales the regularization term.
In contrast the influence of the time-parameter is more subtle as it changes the eigenvalues / eigenvectors of $\Psi$ and thus acts in a nonlinear fashion.

Second, by choosing a particular $\bm{\nu}$ in the projection term such that a time-independent component is picked out from $Y$, we can recover classical some classical null model terms.
As an example we can again consider the symmetric Laplacian dynamics $Y= \exp(-Lt)$, for which the centering operation $\mathcal W_\perp = I - \mathbf{11}^\top /n$ corresponds to projecting out the stationary eigenvector (see also Section~\ref{sec:SI_relationsDiffusion}, where the case $Y=\exp(-D^{-1}Lt)$ and the configuration null model is discussed as well).
Note, however, that in general the second term remains time-dependent and the null model may vary with time, too. 
The choice of a projection may thus be guided either by simple considerations on what aspect of the dynamics we are interested in (e.g., the relative difference of the influence which would lead to a type of centering operation), or by suppressing some type of mode which we know might be irrelevant for our considerations (e.g., the stationary distribution in a diffusion process~\cite{Delvenne2013,Schaub2012}).

\subsection{Optimizing the quality measure \texorpdfstring{$r_\nu(t,H)$} \; using an adapted Louvain algorithm}
The optimization of the quality measure $r_\nu(t,H)$ given in~\eqref{eq:quality} can be achieved by various means, e.g., via MCMC or spectral techniques.
Here we propose to use an augmented version of the Louvain algorithm~\cite{Blondel2008}, which was initially proposed as an efficient algorithm to optimize the Newman-Girvan modularity~\cite{Newman2004}. The algorithm operates as follows:
\begin{enumerate}
        \item Loop over all nodes in a random order, and assign each node greedily to the community for which the increase in quality is maximal until no further move is possible.
        \item Build a coarse-grained network, in which each node represents a community in the previous network.
        \item Repeat steps 1--2, until no further improvement is possible.
\end{enumerate}
As outlined in Ref.~\cite{Campigotto2014}, this generic procedure can be used to optimize any quality function of the form:
\begin{equation}\label{eq:louvain_form}
    \text{trace } H^\top[F- \mathbf{ab}^\top]H,
\end{equation}
where $H$ is the partition indicator matrix, $F$ is a general matrix derived from the network, and $\mathbf{a},\mathbf{b}$ are two $n$ dimensional vectors. 
The quality function~\eqref{eq:quality} is clearly of this form.
An inherent problem of many community detection measures is the choice of the relevant resolution, or scale, of the partitioning.
In many methods this choice has to be made explicitly a priori, by declaring how many groups are to be found by the method. 
If this is not the case, then there there is either a free ('resolution') parameter or a regularization scheme, with which the size of the groups found can be controlled explicitly or implicitly, or there is an implicit scale associated with the method, which will determine an upper and lower limit of size the communities to be found~\cite{Schaub2012,Schaub2012a,Fortunato2007,Lancichinetti2011a}.

Instead of choosing and fixing a particular scale, we here identify significant partitions according to the criteria outlined in Refs.~\cite{Lambiotte2010,Delmotte2011,Schaub2012}. 
We advocate to look at the trajectories for all times $t$ and let thereby the dynamical process reveal the important scales of the problem. These scales should be associated with robust partitions over time and relative to the optimization. 
Thus we are interested in identifying robust partitions as indicated by: (i) a persistence to (small) time-variations, which translates into long plateaux in the number of communities plotted against time; (ii) consistency of partitions obtained from the generalized Louvain algorithm over random initialisation conditions as measured by the mean distance between partitions 
using the normalized Variation of Information (VI) metric~\cite{Meila2007}.  
A VI of zero results when all iterations of the Louvain algorithm return exactly the same clustering.
By computing the matrix $VI(t,t')$, containing the mean variation of information between any two sets of partitions at different times, we can easily identify time-epochs over which we always obtain very similar, robust partitions. 

\subsection{Dynamical roles, modules and symmetries}
\begin{figure}[tb!]
    \centering
    \includegraphics[width=\columnwidth]{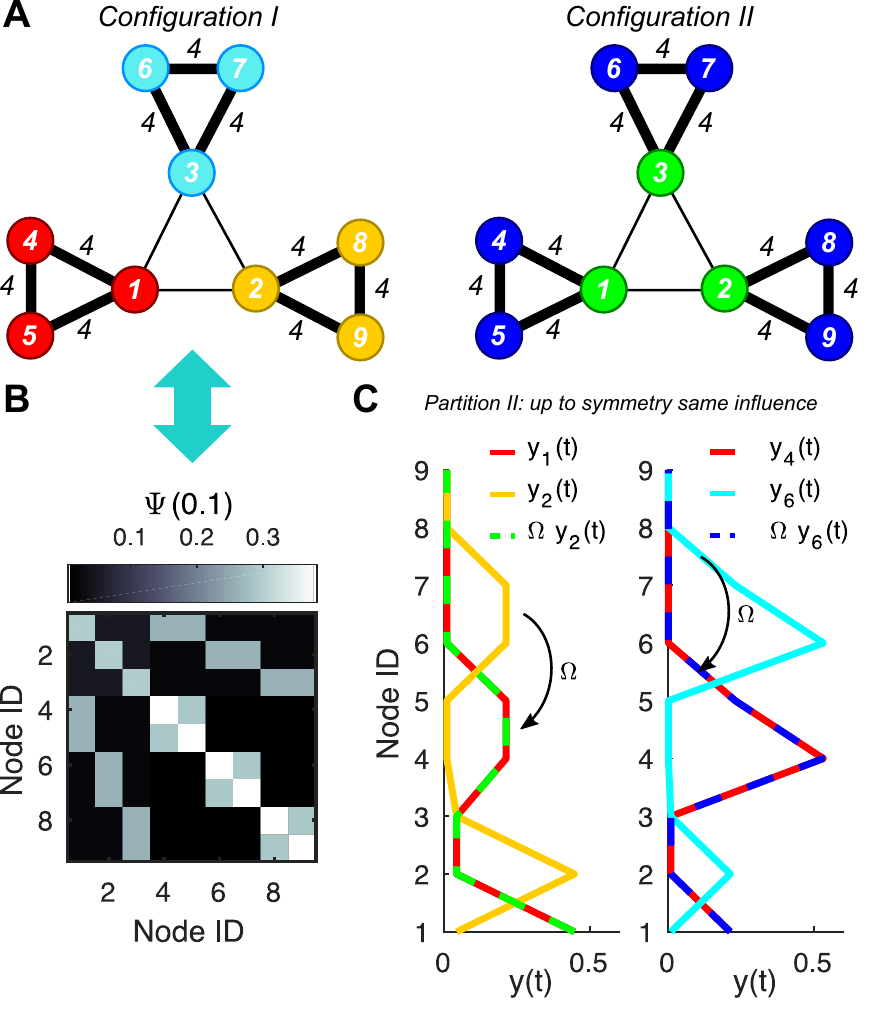}
    \caption{\textbf{Dynamical similarities, modules and roles}. \textbf{A} A small network with 2 possible partitions with nodes may be described as similar. \textbf{B} If we were to group the nodes according to the dynamical similarity measure as defined in the text we would pick out configuration A, as the nodes in each group have a similar impulse response after time $t$.
    \textbf{C} However, if we consider the impulse responses up to the action of a permutation $\Omega$ (isometric mapping), we could also infer the  role' partition of configuration B.}%
    \label{fig:schematic_roles_modules}
\end{figure}

As many notions of dynamical or functional role have been presented in the literature so far, we provide here a short conceptual clarification on what we would consider a `dynamical' module in this work, in the sense that our similarity measures would assign a high similarity score between each node in a module.
To this end consider the example network depicted in Figure~\ref{fig:schematic_roles_modules}.
For simplicity of our exposition we will consider here a simple consensus dynamics of the form $\dot{\mathbf{x}} = -Lx$, where $L$ is the standard graph Laplacian.
In this case it is essentially the structure of the network that dictates how our similarity measures evolves through the spectral properties of the Laplacian.

We consider two possible partitions in Figure~\ref{fig:schematic_roles_modules}A.
Both partition may be seen to correspond to a type of `role' of the nodes in the network.
In this case, as can been seen in Figure~\ref{fig:schematic_roles_modules}B, our notion of similarity is commensurate with partition I, as the impulse responses of the nodes in the colored group influence the same parts of the network in essentially the same way.
Stated differently, if we denote by $\mathbf{y}_i(t)$ the impulse response of node $i$ after time $t$, then after a brief transient period an impulse given to any node in the same group results in approximately the same state vector of the network (e.g. $\mathbf{y}_1(t) \approx\mathbf{y}_{4}(t) \approx\mathbf{y}_{5}(t)$, in case of the red group).

However, the nodes in partition II are indeed similar in the following sense.
If we consider the vectors $\mathbf{y}_i(t)$ up to the action of a symmetry group (permutation), then we can see that indeed partition II groups nodes together which are similar in this sense.
More precisely, call $\Omega$ a permutation matrix corresponding to the orbit partition II indicated in Figure~\ref{fig:schematic_roles_modules}A.
Then for any two nodes $i,j$ in the same group there exist a permutation matrix $\Omega$ such that $\mathbf{y}_i = \Omega\mathbf{y}_j$ (see Figure~\ref{fig:schematic_roles_modules}C) for an illustration. 

One could thus try to search for these kinds of partitions as well from the perspective of our dynamical framework.
We postpone a exploration of these tasks for future work. 
However, see for instance Refs~\cite{Cooper2011,Cooper2010,Faccin2017,Sanchez-Garcia2018,Schaub2016} for related discussions.

\section{The (centered) dynamic similarity \texorpdfstring{$\Psi(t)$} \; for diffusive processes}\label{sec:SI_relationsDiffusion}
In this section we comment on how specific measures can be recovered and extended within the here presented framework, when focussing on diffusion processes.
Note however, as discussed in the main text, the dynamical similarity $\Psi$ is applicable to general linear dynamics (including signed networks). 
Below we consider first the case of a diffusion process on undirected network, before we comment on the diffusion processes on directed networks.

\subsection{The undirected case}\label{sec:undirected}
To put our approach in the context of diffusion processes, let us first consider an undirected dynamics of the form
\begin{equation}\label{eq:lap_diffusion}
  \dot{\mathbf{p}}= -\mathbf{p} \, L, 
\end{equation} 
where $\mathbf{p}$ is the $1\times n$ row vector describing the probability of a particle to be present at any node.
Note that this diffusion is the dual of the consensus process:
\begin{equation}
\label{eq:consensus}
\dot{\mathbf{x}} = -L\mathbf{x},
\end{equation}
and indeed, in this case these two dynamics are in fact identical, as transposing Equation~\eqref{eq:lap_diffusion} corresponds to a dynamics of the form~\eqref{eq:gen_setup} with  $\mathcal{B}=\mathcal{C} =I$ and $\mathcal{A} = - L = -L^\top$, the combinatorial graph Laplacian.

As it is customary in the context of diffusion processes to deal with row vectors, and accordingly many results in the literature are presented in this form we will adopt this convention throughout this section.
All these results can be readily transformed into a column vector setup (or in the directed case, may also be interpreted in the light of the dual consensus process).

Consider the random walk associated with the dynamics~\eqref{eq:lap_diffusion} described by the row indicator vector 
$\mathbf{N}(t) \in {\{0,1\}}^{n}$, where $N_i(t)=1$ if the walker is present at node $i$ at time $t$ and zero otherwise, and the $1 \times n$-dimensional vector $\mathbf{p}(t)$ describes the probability of the walker to be at each node at time $t$.
It is well known that if the process takes places on an undirected (i.e., $L=L^\top$) connected graph, it is guaranteed to be wide sense stationary (in fact ergodic) and $\mathbf{p}(t)$ converges to the unique stationary distribution 
\begin{equation*}
    \bm{\pi} = \frac{\mathbf{1}^\top}{n},
\end{equation*}
irrespective of the initial condition.

To derive further results, it is insightful to compute the auto-covariance matrix of this process.
Let us assume that we prepare the system at stationarity, $\mathbf{p}(0) \sim \bm{\pi}$
(i.e., the walker is equally likely to start at any node at time $t=0$).
The expectation of $\mathbf{N}(t)$ remains constant over time and we have 
\begin{equation*}
    \mathbb E[\mathbf{N}(t)]= \mathbb E[\mathbf{N_0}] \, P(t) =  \boldsymbol \pi P(t) = \boldsymbol \pi,
\end{equation*}
where the transition matrix for this process is given by:
\begin{equation*}
    P(t) = \exp(-Lt).
\end{equation*}
The auto-covariance matrix of the process is then:
\begin{align}\label{eq:cov_autocov}
    \Sigma(t)  &=  \text{cov}\left[{\mathbf{N}(0)}^\top,\mathbf{N}(t)\right]\\
& = \mathbb E[\mathbf{N_0}^\top \mathbf{N}(t)] - \mathbb E [\mathbf{N^\top_0}] \, \mathbb{E} [\mathbf{N_0}] \\
&=\Pi P(t) -\boldsymbol{\pi}^\top\boldsymbol{\pi},
\end{align}
where $\Pi = \text{diag}(\bm{\pi})$.
Defining $\Sigma_0 = \Pi - \boldsymbol{\pi}^\top\boldsymbol{\pi}$, we get that
\begin{align}
\label{eq:autocovariance_diffusion}
\Sigma(t) = \Sigma_0 \, P(t) = \Sigma_0 \, \exp(-Lt),
\end{align}
and it becomes apparent that $\Sigma(t)$ is governed by the matrix differential equation:
\begin{equation}
 \frac{d \Sigma}{dt} = -\Sigma L \quad \text{with} \quad \Sigma(0) = \Sigma_0.
\end{equation}
This autocovariance matrix $\Sigma(t)$ has been used as a dynamic similarity matrix in the Markov Stability framework for community detection~\cite{Delvenne2010,Delvenne2013,Schaub2012}.

Now, let us compare the autocovariance $\Sigma(t)$ with the dynamic similarity $\Psi(t)$.
As shown in~\eqref{eq:lyapunov_xi}, when $\mathcal B=I$ the dynamic similarity $\Psi(t)$ 
obeys a Lyapunov matrix differential equation, which in this diffusive case is:
\begin{flalign}
\label{eq:Lyapunov_eq}
    \frac{d\Psi}{dt} = -L^\top\Psi -\Psi L \quad \text{with} \quad \Psi(0) = \Psi_0.
\end{flalign} 
Without loss of generality we may pick $\Psi_0 = \Sigma_0$ as initial condition, so that
the solution is given by 
\begin{align}
\label{eq:Psi_diffusion}
\Psi(t) = {P(t)}^\top \Sigma_0 P(t) = \exp(-L^{\top}t) \Sigma_0  \exp(-Lt),
\end{align}
which is to be compared to~\eqref{eq:autocovariance_diffusion}

For the case of undirected graphs, we have $L=L^\top$ and $L\mathbf{1}=0$, hence
\begin{equation*}
    \Sigma_0 = \Pi - \boldsymbol{\pi}^\top\boldsymbol{\pi} = \frac{1}{n} \left(I - \frac{\bm{1}\bm{1}^\top}{n} \right).
\end{equation*}
Therefore, we can interpret $\mathcal{W}_\perp= n\Sigma_0$ as a (scaled) projection matrix in Eq.~\eqref{eq:Psi_diffusion}, and~\eqref{eq:Psi_diffusion} as a centered dynamic similarity: 
\begin{align}
    \dfrac{1}{n} \Psi_\perp(t) 
  = \exp(-L t) \frac{1}{n} \left(I - \frac{\bm{1}\bm{1}^\top}{n} \right)  \exp(-L t).
\end{align}

We can then rewrite $\Psi_\perp(t)$ to show the equivalence with $\Sigma(t)$ up to a simple rescaling:
\begin{align*}
    & \frac{1}{n}\Psi_\perp(t) 
  = \frac{1}{n} \exp(-L t) \left(I - \frac{\bm{1}\bm{1}^\top}{n} \right)  \exp(-L t) \nonumber\\
  &= \frac{1}{n} \left(I - \frac{\bm{1}\bm{1}^\top}{n} \right)  \exp(-L t) \exp(-L t)\nonumber\\
  &= \frac{1}{n} \left(I - \frac{\bm{1}\bm{1}^\top}{n} \right)  \exp(-L (2t)) =\Sigma_0 P(2t) = \Sigma(2t),
\end{align*}
Hence for the case of diffusion on undirected graphs, the centered dynamic similarity $\Psi_\perp(t)$ is proportional to the autocovariance of the diffusion on a rescaled time.

Although we have exemplified this connection with a particular example, this result applies to any \textit{time-reversible dynamics}. 
This includes all customary defined diffusion dynamics on undirected graphs like the continuous-time unbiased random walk, the combinatorial Laplacian random walk, or the maximum entropy random walk~\cite{Lambiotte2014}.

This result follows from the reversibility condition for a Markov process~\cite{Bremaud1999,Gallager2013}, also known as detailed balance:
\begin{equation}\label{eq:detailed_balance}
  \pi^{(i)} p_{i\rightarrow j} = \pi^{(j)} p_{j\rightarrow i} \quad \forall \;  i, j,
\end{equation}
i.e., at stationarity, the probability to transition from state $i$ to state $j$ is the same as the probability to transitions from $j$ to $i$ (for any $i,j$).
In matrix terms, the detailed balance condition is:
\begin{equation}
  \Pi P(t) = 
  {P(t)}^\top \Pi,  \quad  \forall t.
\end{equation} 

Let us consider a centered dynamical similarity measure, in which we choose the weighting matrix 
\begin{equation*}
    \mathcal W_\Pi = \Pi -\bm\pi^\top\bm\pi,
\end{equation*}
which can be throught of as a generalization of the standard projection matrix 
$\mathcal{W}_\perp$. 
It is now easy to see that the analogous relationship between $\Psi_\Pi(t)$ and the autocovariance $\Sigma(t)$ holds also under detailed balance:
\begin{align*}
 \Psi_\Pi(t) & = {P(t)}^\top \left(\Pi - {\bm{\pi}}^\top \bm{\pi} \right) \, P(t) \\
 &=  \left( \Pi P(t) - {\bm{\pi}}^\top \bm{\pi} P(t) \right)  \, P(t)  \\
& =  \left(\Pi - {\bm{\pi}}^\top \bm{\pi} \right) \, P(2t) = \Sigma(2t).
\end{align*}

In the case of diffusive dynamics on undirected networks with detailed balance, we have shown that the dynamical similarity $\Psi_\Pi(t)$ is equivalent to the autocovariance $\Sigma(t)$ up to a rescaling. 
Therefore the Louvain-like analysis of $\Psi_\Pi(t)$ on the quality function $r_{\Pi}(t,H) = \text{trace }H^\top\Psi_\Pi(t)H $ can be seen as a proper generalization of the Markov Stability framework, which optimizes the quality function $s(t,H) = \text{trace }H^\top\Sigma(t)H$, and thus encompasses a wide array of notions of community detection including the classical Newman-Girvan Modularity~\cite{Newman2004}, the self-loop adjusted modularity version of Arenas et al.~\cite{Arenas2008}, the Potts model heuristics of Reichardt and Bornholdt~\cite{Reichardt2004}, as well as Traag et al.~\cite{Traag2011}, and classical spectral clustering~\cite{Shi2000}.
For details, we refer the reader to the derivations given in Refs.~\cite{Delvenne2013,Lambiotte2014} in terms of the Markov Stability measure.

Note, however, that Markov Stability only deals with diffusion dynamics, whereas both the centered dynamical similarity $\Psi_\Pi$ (related to the quality function $r_\Pi(t,H)$), and the kernel $\Psi$ can be applied to general linear models (including signed networks), as discussed in the main text.  
An interesting case occurs when considering diffusive processes on directed graphs, as discussed in the next section.

\subsection{The directed case}\label{sec:directed}\label{sec:dirgraphs_and_probs}
\begin{figure}[tb!]
 \centering
 \includegraphics[width =\columnwidth]{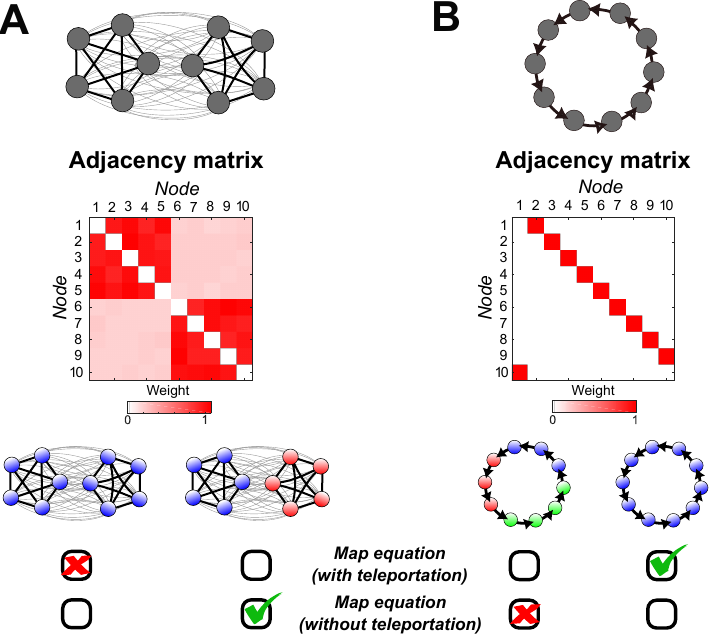}
 \caption{\textbf{Unpredictable influence of teleportation on clustering of directed graphs ($A \neq A^\top$).} 
When studying a directed diffusive dynamics (on a directed graph), a teleportation component is usually added to make the process ergodic. 
The addition of teleportation can lead to unexpected effects when clustering the network, since teleportation also influences the cut (i.e., the probability flow across group boundaries). 
For concreteness, we illustrate this effect here the map-equation~\cite{Rosvall2008} although this effect is general. 
\textbf{A}~A directed network of two cliques. The weights between nodes inside each group are drawn from a uniform distribution with mean $1 \pm 0.1$; the weights across different groups are drawn from a uniform distribution with mean $0.2 \pm 0.02$. 
The network is directional but the asymmetry of $A$ is so weak as to appear visually virtually undirected. 
The introduction of teleportation leads to a resolution limit effect in which the groups cannot be resolved. 
\textbf{B}~A~directed cycle network with equal weights. 
Without teleportation, a split of the ring into multiple groups is found, whereas the introduction of teleportation improves the result in this case so that the whole cycle is detected.
Using an analysis based on $\Psi$ without teleportation~\eqref{eq:Psi_no_teleportation} finds the dynamical blocks directly in both of these examples.
}\label{fig:Teleport}
\end{figure}

For undirected diffusions, the autocovariance  can be used as a dynamic similarity between nodes. However, there are important differences for directed (asymmetric) diffusive processes, as such processes are not guaranteed to be ergodic.

Let us consider a diffusion on a directed graph:
$$\dot{\mathbf{p}} = -\mathbf{p}L$$ 
so that $\mathcal{B}=\mathcal{C}=I$ and $-\mathcal{A}=L  \neq L^\top$, an asymmetric Laplacian. 
This process is only ergodic if the graph is strongly connected. 
In many scenarios this is not the case, however, and the process asymptotically concentrates 
the probability on sink nodes with no outgoing links.
Hence node similarities cannot be based on autocovariances at stationarity.

In order to study node similarities based on the diffusive dynamics, the original process is usually modified by adding a small `teleportation' term (e.g., allowing for the process to diffuse to any node on the graph with a small probability)\cite{Delvenne2010,Rosvall2008}.
This approach is also known as the `Google trick,' as it was popularized through its use in the original computation of Pagerank \cite{Page1999}. In its original form, the introduction of teleportation creates a related strongly connected graph by combining the original graph (with Laplacian $L$) together with the complete graph.  This creates a related (yet different) ergodic process on this surrogate graph which can then be analyzed~\cite{Lambiotte2014} via dynamic similarities based on autocovariances, as discussed in Section~\ref{sec:undirected}. 
Specifically, the surrogate, ergodic system is defined by an adjusted Laplacian operator 
$$\widetilde{L}  = L + L_{\text{teleport}}.$$

Following an analogous calculation as above, the auto-covariance $\Sigma(t)$ of this process:
\begin{align*}
    \widetilde{\Sigma}(t)= \widetilde{\Pi} \exp(-\widetilde{L}  t)- \widetilde{\bm{\pi}}^\top \widetilde{\bm{\pi}} = \left(\widetilde{\Pi} - \widetilde{\bm{\pi}}^\top \widetilde{\bm{\pi}} \right)  \exp(-\widetilde{L}  t),
\end{align*}
where $\widetilde{\bm{\pi}}$ is the stationary distribution of the surrogate ergodic process (e.g., Page Rank). 
However, this autocovariance is now asymmetric, in general, and its interpretation as a similarity matrix is problematic. 

In contrast, we can use the Lyapunov equation~\eqref{eq:Lyapunov_eq} to define the dynamic similarity~\eqref{eq:Psi_diffusion} of the ergodic system  
\begin{align}
\label{eq:Psi_ergodic}
    \widetilde{\Psi}_{\widetilde{\Pi}}(t) = \exp(-\widetilde{L}^\top t) \left(\widetilde{\Pi} - \widetilde{\bm{\pi}}^\top \widetilde{\bm{\pi}} \right)  \exp(-\widetilde{L}  t),
\end{align}
where we have chosen the initial condition $\Psi_0 = \widetilde{\Pi} - \widetilde{\bm{\pi}}^\top \widetilde{\bm{\pi}}$.
Note that Eq.~\eqref{eq:Psi_ergodic} may alternatively be constructed from the dual process $\dot{\mathbf{x}} = \mathcal A \mathbf{x}$, with the operator $\mathcal A = -\widetilde{L}$.
The similarity $\widetilde{\Psi}_{\widetilde{\Pi}}(t)$ may now be exploited directly to carry out embeddings, spectral clusterings, or Louvain-like block detection, as described in the Methods section.  
Note that the analysis based on the dynamic similarity $\widetilde{\Psi}_{\widetilde{\Pi}}(t)$ remains distinct to the symmetrized autocovariance $(\widetilde{\Sigma} + \widetilde{\Sigma}^\top)/2$ which is optimized when using a Louvain-like algorithm~\cite{Lambiotte2014}. 
Other symmetrizations have been introduced in the context of transition matrices in directed graphs~\cite{Satuluri2011} generalizing Kleinberg's HITS scores~\cite{Kleinberg1999}. 

The definition of the dynamic similarity of the associated ergodic process~\eqref{eq:Psi_ergodic} renders it consistent with our generic framework.
However, the introduction of teleportation to create the surrogate process has conceptual disadvantages.

First, teleportation perturbs the dynamics in a non-local manner and induces uncontrolled effects when finding node similarities based on dynamics.
In particular, it can reduce overclustering (a positive effect), but can also lead to (unwanted) resolution limits when finding dynamic blocks (Fig.~\ref{fig:Teleport}).
These issues are only at best mitigated by recent teleportation schemes~\cite{Lambiotte2012}. 

Second, teleportation creates an ergodic, stationary dynamics when key features of the original system might be fundamentally linked to non-stationary data and non-ergodic processes. This can have a bearing on the conclusions drawn from the surrogate ergodic process with added teleportation.   

We illustrate some of these problems in Figure~\ref{fig:Teleport} for the particular example of the map-equation.
However, these issues are generic and effect other diffusion based clustering measures that are based on a notion of persistence of the flow within a region over time, and require a type of ergodicity assumption.
For instance, similar effects will affect the Markov stability measure~\cite{Delvenne2010,Schaub2014}.

Importantly, our dynamic similarity $\Psi(t)$ can be directly applied to non-ergodic, directed graphs without the need to add teleportation (i.e., without creating the associated, but distinct stationary process). In this case, the dynamic similarity is
\begin{align}
\label{eq:Psi_no_teleportation}
\Psi(t) = \exp(-L^{\top}t) \Sigma_0  \exp(-Lt),
\end{align}
which fulfils the Lyapunov equation~\eqref{eq:Lyapunov_eq}. The initial condition $ \Sigma_0$ can be chosen to be any (covariance) matrix which serves as the null model for the process. 
The analysis of this dynamic similarity can reveal dynamic blocks based on directed flows from the \textit{original} diffusive process, as shown in (Fig.~\ref{fig:Teleport}). 
This directed case is another instance where the notion of `dynamic block' generalizes the idea of modules, originally conceived from a structural perspective.

\end{document}